\documentclass[pra,twocolumn,superscriptaddress,showpacs,nofootinbibfloatfix,amsmath,amsfonts,amssymb]{revtex4-1}%
\usepackage{amsmath,amsfonts,amssymb,color}
\usepackage{amsthm}
\usepackage{leftidx}
\usepackage{graphicx}
\usepackage{xcolor}
\usepackage{dcolumn}
\usepackage{bm}
\usepackage{epstopdf}
\usepackage{epsfig}
\usepackage{environ}
\usepackage{pdfcomment}

\usepackage{float}
\usepackage[T1]{fontenc}
\usepackage[latin9]{inputenc}
\usepackage{setspace}
\usepackage{esint}

\begin{document}

\preprint{APS/123-QED}

\title{\textbf{Non-Hermitian Floquet second order topological insulators in periodically quenched lattices}}

\author{Jiaxin Pan}
\affiliation{%
	Department of Physics, College of Information Science and Engineering, Ocean University of China, Qingdao 266100, China
}

\author{Longwen Zhou}
\email{zhoulw13@u.nus.edu}
\affiliation{%
	Department of Physics, College of Information Science and Engineering, Ocean University of China, Qingdao 266100, China
}

\date{\today}

\begin{abstract}
Higher-order topological phases are characterized by protected states
localized at the corners or hinges of the system. By applying time-periodic
quenches to a two-dimensional lattice with balanced gain and loss,
we obtain a rich variety of non-Hermitian Floquet second order topological
insulating phases. Each of the phases is characterized by a pair of
integer topological invariants, which predict the numbers of non-Hermitian Floquet
corner modes at zero and $\pi$ quasienergies. We establish the
topological phase diagram of the model, and find a series of non-Hermiticity
induced transitions between different Floquet second order topological
phases. We further generalize the mean chiral displacement to two-dimensional non-Hermitian systems, and use it to extract the topological
invariants of our model dynamically. This work thus extend the study
of higher-order topological matter to more generic nonequilibrium
settings, in which the interplay between Floquet engineering and non-Hermiticity
yields fascinating new phases.
\end{abstract}

\maketitle

\section{Introduction\label{sec:Intro}}

Higher-order topological phases~(HOTPs) have attracted great attention
in recent years~\cite{HOTP0,HOTP1,HOTP2,HOTP3,HOTP4,HOTP5,HOTP6,HOTP7}. They are featured by localized states appearing at
the boundaries of their boundaries. More precisely, an HOTP of order
$n$~($>1$) in spatial dimension $d$~($\geq n$) possesses topologically protected gapless
states at its $(d-n)$-dimensional boundaries. Over the years, a rich
variety of HOTPs have been found in insulating~\cite{HOTI1,HOTI2,HOTI3,HOTI4,HOTI5,HOTI6,HOTI7,HOTI8,HOTI9,HOTI10,HOTI11,HOTI12,HOTI13,HOTI14,HOTI15,HOTI16,HOTI17,HOTI18,HOTI19,HOTI20}, superconducting~\cite{HOTSC1,HOTSC2,HOTSC3,HOTSC4,HOTSC5,HOTSC6,HOTSC7,HOTSC8,HOTSC9,HOTSC10,HOTSC11,HOTSC12,HOTSC13,HOTSC14,HOTSC15} and
semi-metallic~\cite{HOTSM1,HOTSM2,HOTSM3,HOTSM4,HOTSM5,HOTSM6} systems, and further classified according to their protecting
symmetries~\cite{Classify1,Classify2,Classify3,Classify4}. Experimentally, HOTPs have also been realized in solid-state~\cite{SoliStat1,SoliStat2,SoliStat3,SoliStat4,SoliStat5},
photonic~\cite{Photonic1,Photonic2,Photonic3,Photonic4,Photonic5,Photonic6,Photonic7}, acoustic~\cite{Acoustic1,Acoustic2,Acoustic3,Acoustic4,Acoustic5,Acoustic6,Acoustic7,Acoustic8} and electrical circuit~\cite{Circuit1,Circuit2,Circuit3,Circuit4} platforms, triggering the interest over a wide range of research areas. 

Recently, the study of HOTPs have been extended to nonequilibrium
settings, in which time-periodic driving fields or gains and losses
are applied to a given static system, leading to the discovery of
Floquet HOTPs~\cite{FHOTP1,FHOTP2,FHOTP3,FHOTP4,FHOTP5,FHOTP6,FHOTP7,FHOTP8,FHOTP9,FHOTP10,FHOTP11} and non-Hermitian HOTPs~\cite{NHHOTP1,NHHOTP2,NHHOTP3,NHHOTP4,NHHOTP5,NHHOTP6,NHHOTP7,NHHOTP8}. The Floquet HOTPs are distinguished
from their static cousins by their unique space-time symmetries, topological
invariants, and anomalous Floquet corner or hinge states. On the other
hand, the HOTPs in non-Hermitian systems are featured by non-Bloch
topological invariants, hybrid higher-order skin modes and biorthogonal
bulk-boundary correspondence. Yet, under more general conditions,
a static system could subject to both time-dependent driving fields
and non-Hermitian effects, and much less is known about the fate of
HOTPs in such driven open systems. Moreover, the collaboration of
drivings and dissipation may induce exotic non-Hermitian Floquet HOTPs
that are absent in either closed Floquet systems or non-driven non-Hermitian
systems, which certainly deserve careful investigations. 

In this work, based on the coupled-wire construction of HOTPs~\cite{FHOTP1},
we introduce a class of second order topological insulator (SOTI)
model by coupling an array of one-dimensional (1D) topological insulators
along a second spatial dimension with dimerized hoppings, as presented in Sec.~\ref{sec:Model}. Under the effects of time-periodic
quenches and balanced onsite gains and losses, we find rich non-Hermitian
Floquet SOTI phases in our system, which are protected by the sublattice
and crystal symmetries. In Sec.~\ref{sec:TopInv}, we introduce a pair of integer topological invariants to
characterize the found topological phases, and establish the topological
phase diagram of our model. A series of topological phase transitions
and non-Hermitian Floquet SOTI phases with large topological invariants
are found by varying the amplitude of driving fields or the strength
of gains and losses. Under the open boundary conditions (OBCs), many
non-Hermitian Floquet zero and $\pi$ modes emerge at the corners
of the system, whose numbers are predicted by the bulk topological
invariants, as shown in Sec.~\ref{sec:CorStat}. In Sec.~\ref{sec:MCD}, we propose a way to dynamically extract the topological invariants and detect the topological phase transitions of our system by measuring the mean chiral displacements of a wave packet. Finally, we summarize our results and discuss the possible experimental realizations of our model in Sec.~\ref{sec:Summary}.

\section{Model and symmetry\label{sec:Model}}
In this section, we first introduce an SOTI model following the coupled-wire
construction of static and Floquet SOTIs~\cite{FHOTP1}. Our non-Hermitian
Floquet SOTI system is then realized by applying time-periodic quenches
and balanced onsite gains and losses to the static SOTI model.

\begin{figure}
	\begin{centering}
		\includegraphics[scale=0.4]{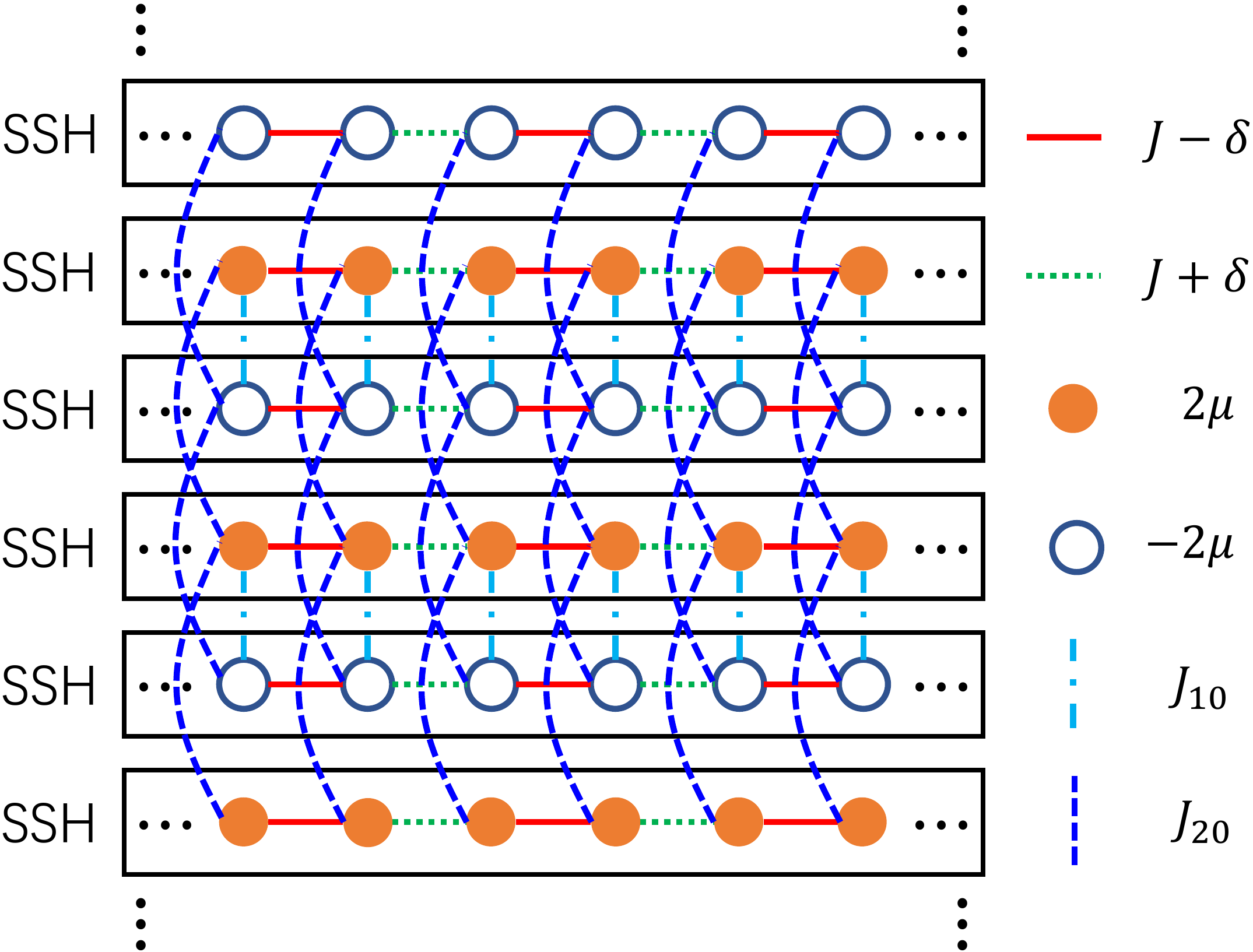}
		\par\end{centering}
	\caption{The schematic diagram of the lattice model described by Eq.~(\ref{eq:HL}). An array of SSH chains are stacked along the vertical ($y$) direction, coupled with each other by the hopping amplitudes $J_{10}$ and $J_{20}$, and also subject to an onsite potential bias $\pm2\mu$.\label{fig:Sketch}}
\end{figure}

We start with a prototypical tight-binding Hamiltonian $H$, which
describes particles hopping on a two-dimensional (2D) square lattice,
\begin{alignat}{1}
H= & \sum_{i,j}[J+(-1)^{i}\delta](|i,j\rangle\langle i+1,j|+{\rm H.c.})\nonumber \\
+ & \sum_{i,j}J_{10}(|i,2j\rangle\langle i,2j+1|+{\rm H.c.})\label{eq:HL}\\
+ & \sum_{i,j}(-1)^{j}(iJ_{20}|i,j\rangle\langle i,j+2|-\mu|i,j\rangle\langle i,j|+{\rm H.c.}).\nonumber
\end{alignat}
Here $i$ ($j$) denotes the lattice site index along the $x$ ($y$)
direction of the system.
An illustration of the lattice model is presented in Fig.~\ref{fig:Sketch}.
Along the $x$-direction, $J-\delta$ ($J+\delta$)
corresponds to the intracell (intercell) hopping amplitude. Along
the $y$-direction, $J_{10}$ and $J_{20}$ characterize the nearest-
and next-nearest-neighbor hopping amplitudes, and $\mu$ denotes the
strength of a staggered onsite potential. The system described by
$H$ can thus be viewed as an array of tight-binding wires lying along
the $y$-direction, with each of them being connected to its adjacent
neighbors by Su-Schrieffer-Heeger (SSH)-type dimerized couplings~\cite{AsbothBook}. Such
kind of ``coupled-wire construction'' has been demonstrated to be
a powerful way of engineering both static and Floquet SOTIs~\cite{FHOTP1} in closed systems.
Generally speaking, four zero-energy topological corner modes would appear
in the system described by Eq.~(\ref{eq:HL}) if both the SSH-type couplings along the $x$-direction
and the wires along the $y$-direction are set in topologically nontrivial
regimes.

Taking periodic boundary conditions (PBCs) along both $x,y$ directions
and performing Fourier transformations, we can express Eq.~(\ref{eq:HL}) in the momentum
representation as $H=\sum_{k_{x},k_{y}}|k_{x},k_{y}\rangle H(k_{x},k_{y})\langle k_{x},k_{y}|$,
where the Hamiltonian matrix $H(k_{x},k_{y})$ has a Kronecker sum
structure
\begin{equation}
H(k_{x},k_{y})=H_{x}(k_{x})\otimes\tau_{0}+\sigma_{0}\otimes H_{y}(k_{y}),\label{eq:Hkxky}
\end{equation}
with
\begin{alignat}{1}
H_{x}(k_{x})& = [(J-\delta)+(J+\delta)\cos k_{x}]\sigma_{x}+(J+\delta)\sin k_{x}\sigma_{y},\label{eq:Hxkx}\\
H_{y}(k_{y})& = 2J_{10}\cos k_{y}\tau_{x}+2(\mu+J_{20}\sin k_{y})\tau_{z}.\label{eq:Hyky}
\end{alignat}
Here $k_{x},k_{y}\in[-\pi,\pi)$ are the quasimomenta along $x$ and
$y$ directions. $\sigma_{0}$ and $\tau_{0}$ are both $2\times2$
identity matrices, with $\mathbb{I}_{4}\equiv\sigma_{0}\otimes\tau_{0}$.
$\sigma_{x,y,z}$ and $\tau_{x,y,z}$ are Pauli matrices acting in
the sublattice spaces in the $x$ and $y$ directions, respectively.
It is well known that both $H_{x}(k_{x})$ and $H_{y}(k_{y})$ describe
1D topological insulators, which are characterized by integer winding
numbers~\cite{AsbothBook}. When both the conditions $|J-\delta|<|J+\delta|$ and $|\mu|<|J_{20}|$ are satisfied,
the 1D descendant systems $H_{x}(k_{x})$ and $H_{y}(k_{y})$ are both in
topologically nontrivial phases at half-filling. In this case, according
to the analysis in Ref.~\cite{FHOTP1}, the parent Hamiltonian $H$ describes
an SOTI with four corner modes under the OBCs. 

In this work, we investigate whether the interplay between time-periodic
drivings and dissipation effects could induce exotic non-Hermitian
Floquet SOTI phases with multiple topological corner modes in the
system described by $H$. To do so, we introduce balanced gain and
loss to the staggered onsite potential $\mu$, i.e., by setting $\mu=u+{\rm i}v$
with $u,v\in\mathbb{R}$. Furthermore, we apply piecewise time-periodic
quenches to each of the wires along the $y$-direction, so that $H_{y}(k_{y})$
becomes
\begin{equation}
H_{y}(k_{y},t)=\begin{cases}
2J_{1}\cos k_{y}\tau_{x} & t\in[\ell T,\ell T+T/2)\\
2(\mu+J_{2}\sin k_{y})\tau_{z} & t\in[\ell T+T/2,\ell T+T)
\end{cases},\label{eq:Hykyt}
\end{equation}
where $t$ is time, $T$ is the driving period and $\ell\in \mathbb{Z}$ counts the number of driving periods. The form of $H_x(k_x)$ remains to be the same during the whole driving period. It is clear that with the driving fields, the hopping amplitude $J_1$ is only turned on in the first half of a driving period. In the second half of the period, the onsite potential $\mu$ and hopping amplitude $J_2$ are switched on. Since the parameters $\mu,J_1$ and $J_2$ only couple intracell degrees of freedom and nearest-neighbor unit cells, the periodic quenches of these parameters are expected to be achievable in recent cold atom~\cite{PQRealize1} and photonic~\cite{PQRealize2} experimental setups. Moreover, as will be made clear in the following sections, the choice of our quench protocol allows the system to close and reopen its spectral gaps alternatively at the quasienergies zero and $\pi$ with the change of system parameters. This has also been demonstrated before in the study of Hermitian Floquet SOTIs~\cite{FHOTP1}. Our system could thus possess rich non-Hermitian Floquet SOTI phases, multiple topological phase transitions and many Floquet corner modes following the choice of our quench protocol.

With Eqs.~(\ref{eq:Hkxky}) and (\ref{eq:Hykyt}), the time-dependent Hamiltonian of our periodically quenched system can be expressed as $H(k_x,k_y,t)=H_x(k_x)\otimes\tau_0+\sigma_0\otimes H_y(k_y,t)$. The resulting Floquet operator, which generates the time evolution of the system over a complete driving period $T$, is then given by $U=\sum_{k_{x},k_{y}}|k_{x},k_{y}\rangle U(k_{x},k_{y})\langle k_{x},k_{y}|$, with $U(k_x,k_y)={\cal {\hat T}}e^{-{\rm i}\int^1_0 dt H(k_x,k_y,t)}$. Here ${\cal {\hat T}}$ performs the time ordering, and we have set the unit of energy to be $\hbar/T$, with $\hbar=T=1$. Since the Hamiltonian of the system stays the same within the first and second halves of the driving period, the integration over time on the exponential of $U(k_x,k_y)$ can be worked out analytically, leading to
\begin{alignat}{1}
U(k_{x},k_{y})& = e^{-\frac{{\rm i}}{2}[H_{x}(k_{x})\otimes\tau_{0}+2(\mu+J_{2}\sin k_{y})\sigma_{0}\otimes\tau_{z}]}\nonumber \\
& \times e^{-\frac{{\rm i}}{2}[H_{x}(k_{x})\otimes\tau_{0}+2J_{1}\cos k_{y}\sigma_{0}\otimes\tau_{x}]}.
\end{alignat}
Noting that $H_{x}(k_{x})\otimes\tau_{0}$ commutes with $2J_{1}\cos k_{y}\sigma_0\otimes\tau_{x}$ and $2(\mu+J_{2}\sin k_{y})\sigma_0\otimes\tau_{z}$, the expression for $U(k_{x},k_{y})$ can be further simplified to
\begin{alignat}{1}
U(k_{x},k_{y}) & =e^{-{\rm i}H_{x}(k_{x})\otimes\tau_{0}}e^{-{\rm i}(\mu+J_{2}\sin k_{y})\sigma_{0}\otimes\tau_{z}}\nonumber\\
& \times e^{-{\rm i}J_{1}\cos k_{y}\sigma_{0}\otimes\tau_{x}}.\label{eq:UkxkyTmp}
\end{alignat}
Finally, expanding each term on the right hand side of $U(k_{x},k_{y})$ into a Taylor series, and combining the relevant terms, we obtain
\begin{alignat}{1}
e^{-{\rm i}H_{x}(k_{x})\otimes\tau_{0}} & =\sum_{n=0}^{\infty}\frac{[-{\rm i}H_{x}(k_{x})]^{n}}{n!}\otimes\tau_{0}\nonumber \\
& =e^{-{\rm i}H_{x}(k_{x})}\otimes\tau_{0},\\
e^{-{\rm i}(\mu+J_{2}\sin k_{y})\sigma_{0}\otimes\tau_{z}} & =\sigma_{0}\otimes\sum_{n=0}^{\infty}\frac{[-{\rm i}(\mu+J_{2}\sin k_{y})\tau_{z}]^{n}}{n!}\nonumber \\
& =\sigma_{0}\otimes e^{-{\rm i}(\mu+J_{2}\sin k_{y})\tau_{z}},\\
e^{-{\rm i}J_{1}\cos k_{y}\sigma_{0}\otimes\tau_{x}} & =\sigma_{0}\otimes\sum_{n=0}^{\infty}\frac{(-{\rm i}J_{1}\cos k_{y}\tau_{x})^{n}}{n!}\nonumber \\
& =\sigma_{0}\otimes e^{-{\rm i}J_{1}\cos k_{y}\tau_{x}}.
\end{alignat}
Plugging these three terms into the right hand side of Eq.~(\ref{eq:UkxkyTmp}), we arrive at
\begin{equation}
U(k_{x},k_{y})=e^{-{\rm i}H_{x}(k_{x})}\otimes e^{-{\rm i}(\mu+J_{2}\sin k_{y})\tau_{z}}e^{-{\rm i}J_{1}\cos k_{y}\tau_{x}},\label{eq:Ukxky}
\end{equation}
which gives the Floquet operator of our system at a fixed quasimomentum $(k_x,k_y)$. Without loss of generality, we choose to work within the topological flat-band limit of $H_{x}(k_{x})$, which can be achieved by setting $J=\delta=\Delta/2$~\cite{AsbothBook}. Experimentally, such an SSH Hamiltonian can be realized in cold atom systems~\cite{SSHRealize1}. With these considerations, the Floquet operator of our system further simplifies to
\begin{equation}
U(k_{x},k_{y})=e^{-{\rm i}H_{0}(k_{x})}\otimes e^{-{\rm i}h_{z}(k_{y})\tau_{z}}e^{-{\rm i}h_{x}(k_{y})\tau_{x}},\label{eq:Ukxy}
\end{equation}
where
\begin{alignat}{1}
H_{0}(k_{x}) & =\Delta(\cos k_{x}\sigma_{x}+\sin k_{x}\sigma_{y}),\label{eq:H0}\\
h_{x}(k_{y}) & =J_{1}\cos k_{y},\label{eq:hx}\\
h_{z}(k_{y}) & =u+{\rm i}v+J_{2}\sin k_{y}.\label{eq:hz}
\end{alignat}
Note that $U(k_{x},k_{y})$ is nonunitary due to the balanced gain
and loss terms $\pm{\rm i}v$ in the staggered onsite potential $\mu\tau_{z}$.
In cold-atom systems, this non-Hermitian onsite potential maybe realized by kicking the atoms out of a trap by a resonant optical beam~\cite{NHRealize1}, or applying a radio-frequency pulse to excite atoms to an irrelevant state, in which an antitrap is further applied to induce the losses~\cite{NHRealize2}.

Before characterizing the topological properties of our non-Hermitian
Floquet system, we first analyze the symmetries that allow it to possess
corner modes at zero and $\pi$ quasienergies under the OBCs. Following
the established approach to the symmetry analysis of Floquet operators~\cite{AsbothSTF,FloquetRev1}, we first transform $U(k_{x},k_{y})$ in Eq.~(\ref{eq:Ukxy}) to a pair of symmetric
time frames upon similarity transformations, yielding
\begin{equation}
U_{\alpha}(k_{x},k_{y})={\cal U}_{0}(k_{x})\otimes{\cal U}_{\alpha}(k_{y}).\label{eq:Uakxy}
\end{equation}
Here $\alpha=1,2$ and 
\begin{alignat}{1}
{\cal U}_{0}(k_{x})& = e^{-{\rm i}H_{0}(k_{x})},\label{eq:U0kx}\\
{\cal U}_{1}(k_{y})& = e^{-{\rm i}\frac{h_{x}(k_{y})}{2}\tau_{x}}e^{-{\rm i}h_{z}(k_{y})\tau_{z}}e^{-{\rm i}\frac{h_{x}(k_{y})}{2}\tau_{x}},\label{eq:U1ky}\\
{\cal U}_{2}(k_{y})& = e^{-{\rm i}\frac{h_{z}(k_{y})}{2}\tau_{z}}e^{-{\rm i}h_{x}(k_{y})\tau_{x}}e^{-{\rm i}\frac{h_{z}(k_{y})}{2}\tau_{z}}.\label{eq:U2ky}
\end{alignat}
It is clear that $U(k_{x},k_{y})$, $U_{1}(k_{x},k_{y})$ and $U_{2}(k_{x},k_{y})$
are similar to one another, and therefore sharing the same complex
Floquet quasienergy spectrum. Furthermore, both $U_{1}(k_{x},k_{y})$
and $U_{2}(k_{x},k_{y})$ possess the sublattice symmetry ${\cal S}=\sigma_{z}\otimes\tau_{y}$,
i.e.,
\begin{equation}
{\cal S}U_{\alpha}(k_{x},k_{y}){\cal S}=U_{\alpha}^{-1}(k_{x},k_{y})
\end{equation}
for $\alpha=1,2$, with ${\cal S}={\cal S}^{\dagger}$ and ${\cal S}^{2}={\mathbb I}_4$.
Besides, we can also identify the diagonal (${\cal M}_{+}$) and off-diagonal
(${\cal M}_{-}$) spatial symmetries of $U_{\alpha}$, i.e.,
\begin{equation}
{\cal M}_{\pm}U_{\alpha}(k_{x}=\pm k_{y}){\cal M}_{\pm}^{-1}=U_{\alpha}^{-1}(k_{x}=\pm k_{y}).
\end{equation}
The spatial symmetries ${\cal M}_{\pm}$ (which happen to be equal to
${\cal S}$ here) guarantee the zero- and $\pi$-quasienergy Floquet
topological modes, if presence, should appear at the four corners of the system
under the OBCs, whereas the topological degeneracy of these non-Hermitian Floquet
corner modes are protected by the sublattice symmetry ${\cal S}$~\cite{FHOTP1}. The sublattice symmetry ${\cal S}$ allows us to introduce a pair of integer winding numbers to characterize the topological phases of our system, as will be discussed in the following section.

\section{Topological invariants\label{sec:TopInv}}

With the relevant symmetries ${\cal S}$ and ${\cal M}_{\pm}$ being identified, we will now introduce the topological invariants of our system.

According to the topological classification of Floquet operators~\cite{AsbothSTF,FloquetRev1},
a Floquet system in one-dimension is characterized by integer winding
numbers. This has been demonstrated for both Hermitian~\cite{DerekPRB,LWZH1,LWZH2,LWZH3} and non-Hermitian~\cite{LWZNH1,LWZNH2,LWZNH3,LWZNH4,LWZNH5}
Floquet models. Since the Floquet operator $U(k_{x},k_{y})$ of our
system in Eq.~(\ref{eq:Ukxy}) has a Kronecker product structure,
its topological invariants may be constructed from the winding numbers
of descendant 1D models ${\cal U}_{0}(k_{x})$ and ${\cal U}_{\alpha}(k_{y})$
($\alpha=1,2$) in the symmetric time frames. To do so, we first note
that ${\cal U}_{0}(k_{x})$ is simply the evolution operator of a
static SSH model over a period. Its topological winding number $w$ is therefore
equal to $1$ ($0$) in the topologically nontrivial (trivial) regime~\cite{AsbothBook}.
For the parameter choice in Eq.~(\ref{eq:H0}), we simply have $w=1$. Furthermore, applying the Euler formula to Eqs.~(\ref{eq:U1ky})
and (\ref{eq:U2ky}), the Floquet operator ${\cal U}_{\alpha}(k_{y})$
($\alpha=1,2$) can be expanded as
\begin{equation}
{\cal U}_{\alpha}(k_{y})=\cos\left[{\cal E}(k_{y})\right]-i\left[d_{\alpha x}(k_{y})\tau_{x}+d_{\alpha z}(k_{y})\tau_{z}\right].
\end{equation}
Here the complex quasienergy dispersion
\begin{equation}
{\cal E}(k_{y})=\arccos\left\{ \cos\left[h_{x}(k_{y})\right]\cos\left[h_{z}(k_{y})\right]\right\} ,\label{eq:Eky}
\end{equation}
and the components of the complex-valued vectors $\left[d_{\alpha x}(k_{y}),d_{\alpha z}(k_{y})\right]$ for $\alpha=1,2$ are given by
\begin{alignat}{1}
d_{1x}(k_{y})& = \sin[h_{x}(k_{y})]\cos[h_{z}(k_{y})],\label{eq:d1x}\\
d_{1z}(k_{y})& = \sin[h_{z}(k_{y})],\label{eq:d1z}\\
d_{2x}(k_{y})& = \sin[h_{x}(k_{y})],\label{eq:d2x}\\
d_{2z}(k_{y})& = \cos[h_{x}(k_{y})]\sin[h_{z}(k_{y})].\label{eq:d2z}
\end{alignat}
The number of times that the two-component vector $\left[d_{\alpha x}(k_{y}),d_{\alpha z}(k_{y})\right]$
winds around zero when $k_{y}$ sweeps across the first Brillouin
zone defines the topological winding number of ${\cal U}_{\alpha}(k_{y})$ for $\alpha=1,2$~\cite{LWZNH1,LWZNH2,LWZNH3,LWZNH4}, i.e.,
\begin{equation}
w_{\alpha}=\int_{-\pi}^{\pi}\frac{dk_{y}}{2\pi}\frac{d_{\alpha x}\partial_{k_{y}}d_{\alpha z}-d_{\alpha z}\partial_{k_{y}}d_{\alpha x}}{d_{\alpha x}^{2}+d_{\alpha z}^{2}}.\label{eq:w12}
\end{equation}
Note in passing that even though $d_{\alpha x}(k_{y})$ and $d_{\alpha z}(k_{y})$
can take complex values, their imaginary parts would have no contributions
to $w_{\alpha}$ as shown in Ref.~\cite{LWZNH4}.

Using the winding numbers $w$ and $w_{\alpha}$ defined separately
for ${\cal U}_{0}(k_{x})$ and ${\cal U}_{\alpha}(k_{y})$ ($\alpha=1,2$) in Eq.~(\ref{eq:Uakxy}),
we can construct the topological invariants of the 2D Floquet operator
$U_{\alpha}(k_{x},k_{y})$ in the $\alpha$'s time frame as
\begin{equation}
\nu_{\alpha}=ww_{\alpha},\qquad\alpha=1,2.\label{eq:nu12}
\end{equation}
Since the Floquet operator $U(k_x,k_y)$ could open gaps at both the quasienergies zero and $\pi$, we need at least two invariants to characterize its topological phases.
By combining $\nu_{1}$ and $\nu_{2}$, we obtain such a pair
of integer topological invariants $(\nu_0,\nu_{\pi})$, given by
\begin{equation}
\nu_{0}=\frac{\nu_{1}+\nu_{2}}{2},\qquad\nu_{\pi}=\frac{\nu_{1}-\nu_{2}}{2}.\label{eq:nu0P}
\end{equation}
It tends out that the invariants $(\nu_{0},\nu_{\pi})$
could fully characterize the non-Hermitian Floquet SOTI phases of
$U(k_{x},k_{y})$ in Eq.~(\ref{eq:Ukxky}). 
They take quantized values so long as the sublattice symmetry ${\cal S}$ is preserved.
Moreover, we will demonstrate that under the OBCs,
these topological invariants correctly predict the numbers of non-Hermitian
Floquet corner modes at zero and $\pi$ quasienergies, thus establishing the bulk-corner correspondence of our system. In the Hermitian
limit, the invariants $(\nu_{0},\nu_{\pi})$ could also characterize
the SOTI phases of the resulting closed Floquet system~\cite{FHOTP1}, so long as the corresponding
Floquet operator $U(k_{x},k_{y})$ shares the same tensor product structure with Eq.~(\ref{eq:Ukxky}).

\section{Topological phase diagram\label{sec:PhsDiag}}

In this section, based on the topological invariants introduced in
Eq.~(\ref{eq:nu0P}), we present the topological phase diagram of
our non-Hermitian Floquet SOTI model in typical situations.

From Eqs.~(\ref{eq:w12}) and (\ref{eq:nu12}), it is clear that $\nu_{0}\neq0$
($\nu_{\pi}\neq0$) in Eq.~(\ref{eq:nu0P}) if both $w$ and $\frac{w_{1}+w_{2}}{2}$
($\frac{w_{1}-w_{2}}{2}$) are nonzero. As the parameters of the 1D descendant
system in the $x$-direction in Eq.~(\ref{eq:H0}) has been set inside
the topological nontrivial regime, we have the winding number $w=1$
for ${\cal U}_{0}(k_{x})$. A topological phase transition in our
system is then accompanied by the closing and reopening of a spectral
gap of ${\cal U}_{\alpha}(k_{y})$ at the quasienergy zero or $\pi$
on the complex plane. 

According to Eq.~(\ref{eq:Eky}), the gapless condition of ${\cal U}_{\alpha}(k_{y})$
is determined by
\begin{equation}
\cos\left[{\cal E}(k_{y})\right]=\cos\left[h_{x}(k_{y})\right]\cos\left[h_{z}(k_{y})\right]=\pm1,\label{eq:EkyGapless}
\end{equation}
where the $+1$ ($-1$) on the right hand side of Eq.~(\ref{eq:EkyGapless}) corresponds to a gap closing at ${\cal E}(k_{y})=0$ {[}${\cal E}(k_{y})=\pi${]}. With
the help of Eqs.~(\ref{eq:hx}) and (\ref{eq:hz}), Eq.~(\ref{eq:EkyGapless})
is equivalent to the following two equalities
\begin{alignat}{1}
\sin(u+J_{2}\sin k_{y}) & =0,\label{eq:PhsBd1}\\
\cos(J_{1}\cos k_{y})\cos(u+J_{2}\sin k_{y})\cosh v & =\pm1.\label{eq:PhsBd2}
\end{alignat}
Combining them together, we can express the gapless condition of
${\cal U}_{\alpha}(k_{y})$ in Eq.~(\ref{eq:Uakxy}) as
\begin{equation}
v=\pm{\rm arccosh}\left\{ \frac{1}{\cos\left[J_{1}\sqrt{1-(n\pi-u)^{2}/J_{2}^{2}}\right]}\right\} ,\label{eq:PhsBd}
\end{equation}
where $n\in\mathbb{Z}$ and $|n\pi-u|<|J_{2}|$. Eq.~(\ref{eq:PhsBd})
determines the boundaries between different topological phases in
the parameter space, across which the system described by $U(k_{x},k_{y})$
in Eq.~(\ref{eq:Ukxy}) is expected to change from one non-Hermitian Floquet SOTI phase to
another.

\begin{figure}
	\begin{centering}
		\includegraphics[scale=0.49]{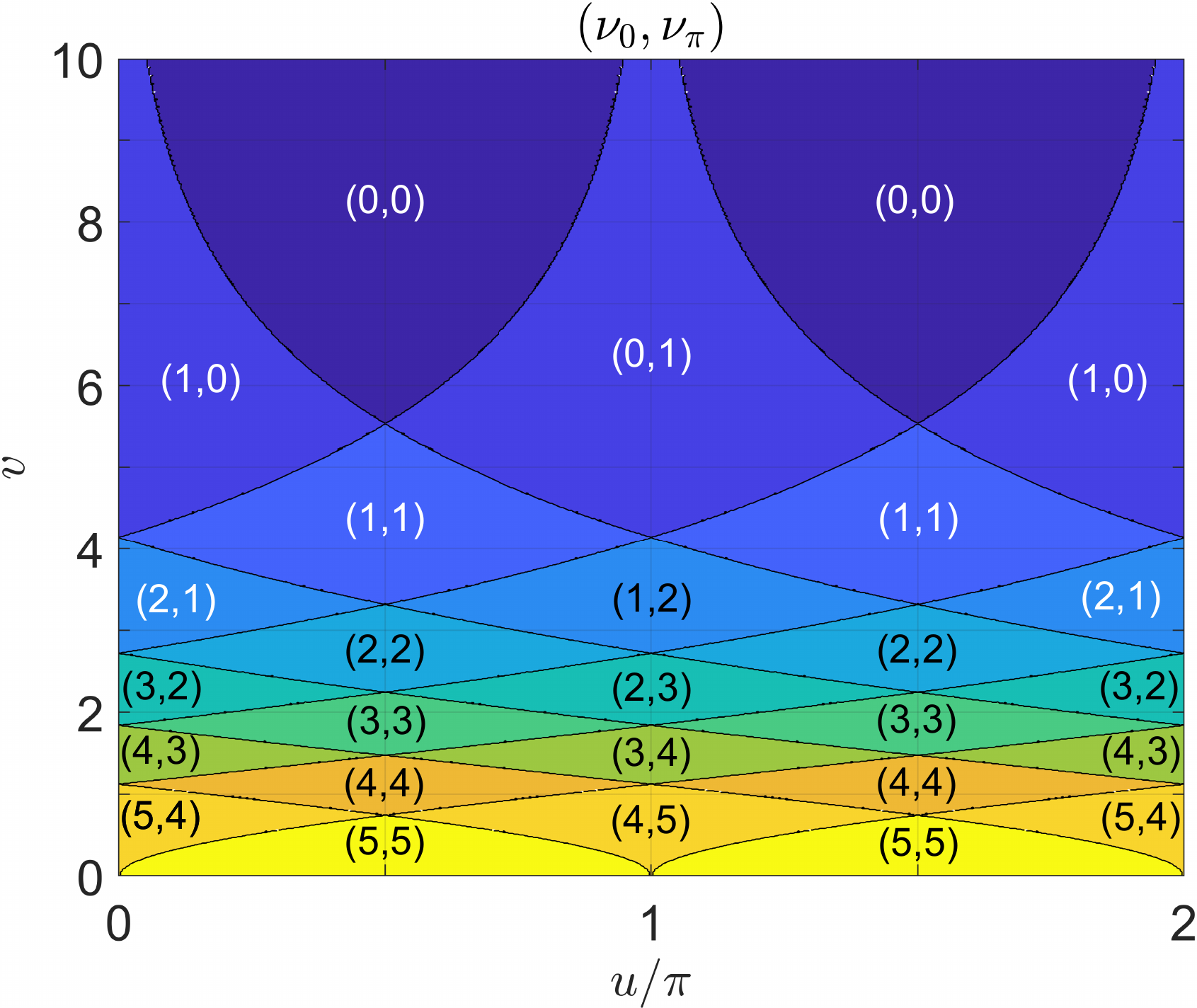}
		\par\end{centering}
	\caption{Topological phase diagram of the periodically quenched non-Hermitian
		lattice model (\ref{eq:Ukxy}) versus the real and imaginary parts
		of onsite potential $\mu=u+{\rm i}v$. The other system parameters
		are fixed at $\Delta=\pi/20$, $J_{1}=0.5\pi$ and $J_{2}=5\pi$.
		Each region with a uniform color corresponds to a non-Hermitian Floquet
		SOTI phase, whose topological invariants $(\nu_{0},\nu_{\pi})$ are
		denoted explicitly therein. The black lines separating different phases
		are determined by the gapless condition Eq.~(\ref{eq:PhsBd}).\label{fig:PhsDiagUV}}
\end{figure}

In the following, we present the topological phase diagrams of our
periodically quenched non-Hermitian lattice model Eq.~(\ref{eq:Ukxky}) in three typical situations.
In the first case, we show the phase diagram versus the real and imaginary
parts of the onsite potential $\mu=u+{\rm i}v$ in Fig.~\ref{fig:PhsDiagUV}.
The other system parameters are chosen as $J=\delta=\Delta/2=\pi/40$,
$J_{1}=0.5\pi$ and $J_{2}=5\pi$. The values of topological invariants
$(\nu_{0},\nu_{\pi})$, obtained from Eqs.~(\ref{eq:w12})-(\ref{eq:nu0P}),
are shown explicitly in Fig.~\ref{fig:PhsDiagUV} within each of the
non-Hermitian Floquet SOTI phases. The black lines separating different
phases (regions with different colors) in Fig.~\ref{fig:PhsDiagUV}
are obtained from the gapless condition Eq.~(\ref{eq:PhsBd}). From the phase diagram, we
observe a series of topological phase transitions accompanied by quantized
jumps of $\nu_{0}$ and/or $\nu_{\pi}$ by varying either the real
or imaginary part of $\mu$. Therefore, the existence of balanced onsite
gains and losses can indeed induce phase transitions and new types of
non-Hermitian Floquet SOTIs in our system. Furthermore, we found a
couple of SOTI phases characterized by large topological invariants
$(\nu_{0},\nu_{\pi})$. Detailed numerical calculations suggest that
the values of $(\nu_{0},\nu_{\pi})$ can be arbitrarily large with
the increase of the hopping amplitude $J_{2}$. These SOTI phases originate
from the interplay between the time-periodic driving fields and the
onsite gains and losses. They are thus unique to non-Hermitian Floquet
systems. Under the OBCs, a non-Hermitian Floquet SOTI phase with large
invariants $(\nu_{0},\nu_{\pi})$ will also admit multiple quartets
of topological corner modes at zero and $\pi$ quasienergies, as will
be demonstrated in the next section.

\begin{figure}
	\begin{centering}
		\includegraphics[scale=0.49]{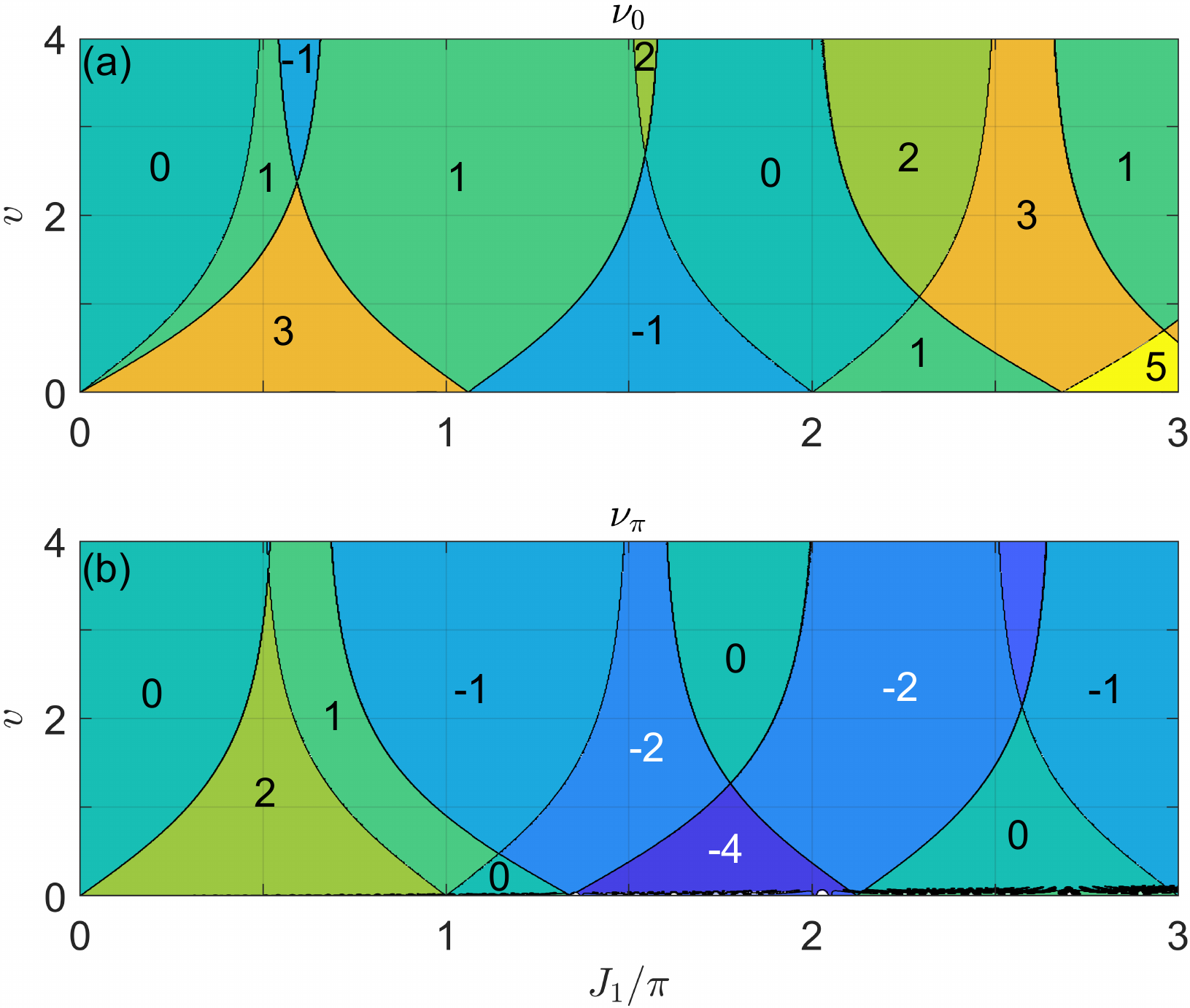}
		\par\end{centering}
	\caption{Topological phase diagram of the periodically quenched non-Hermitian
		lattice model (\ref{eq:Ukxy}) versus the hopping amplitude $J_{1}$
		and gain or loss amplitude $v$. The other system parameters are
		$\Delta=\pi/20$, $J_{2}=3\pi$ and $u=0$. The values of topological invariants $\nu_{0}$ ($\nu_{\pi}$) for each non-Hermitian Floquet SOTI
		phase with a uniform color are shown in panel (a) {[}(b){]}. The black
		lines separating different phases are obtained from the gapless condition
		Eq.~(\ref{eq:PhsBd}).\label{fig:PhsDiagJ1V}}
\end{figure}
\begin{figure}
	\begin{centering}
		\includegraphics[scale=0.5]{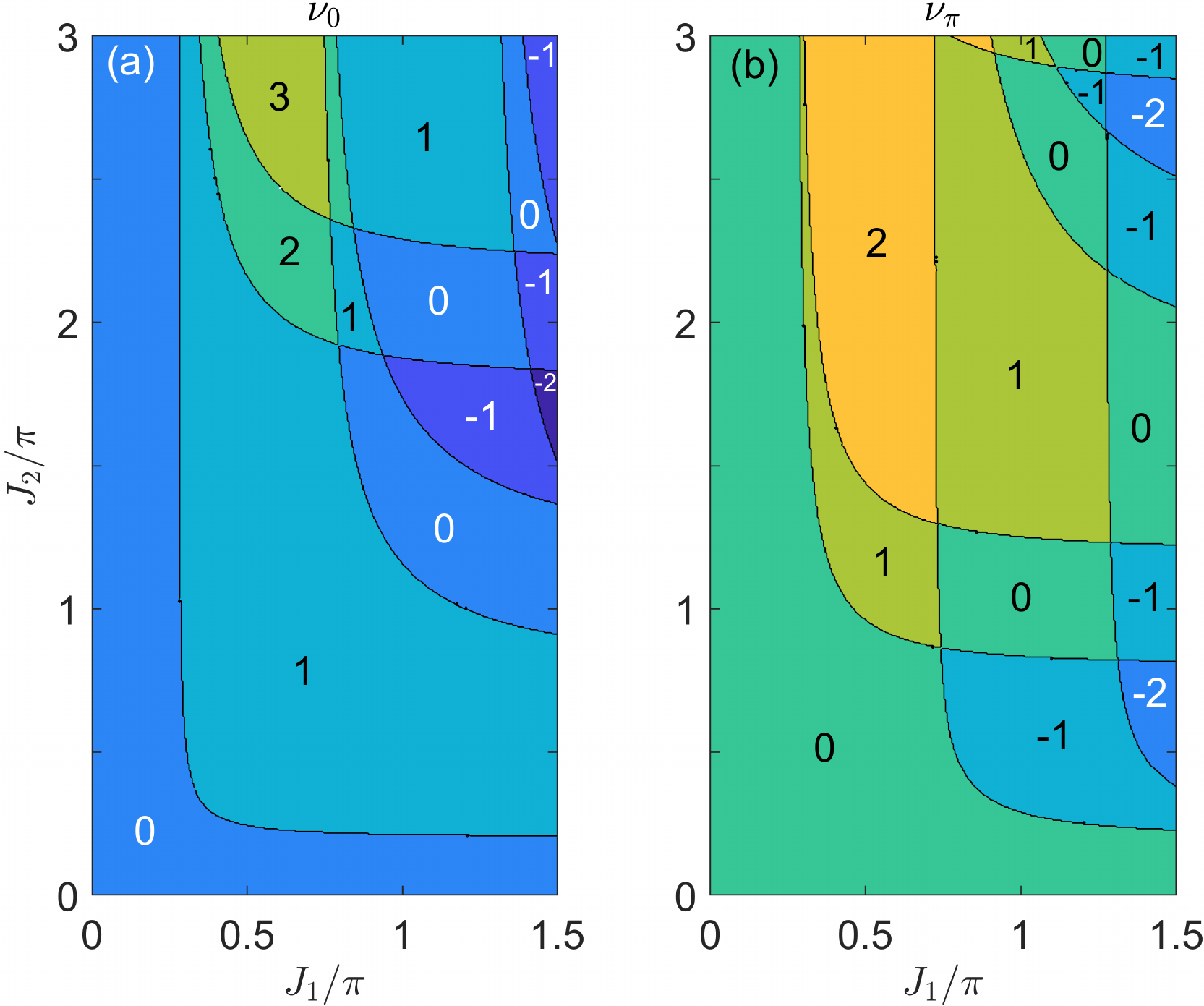}
		\par\end{centering}
	\caption{Topological phase diagram of the periodically quenched non-Hermitian
		lattice model (\ref{eq:Ukxy}) versus the hopping amplitudes $J_{1}$
		and $J_{2}$. The other system parameters are
		$\Delta=\pi/20$, $u=0.2\pi$ and $v=1{\rm i}$. The values of topological invariants $\nu_{0}$ ($\nu_{\pi}$) for each non-Hermitian Floquet SOTI
		phase with a uniform color are shown in panel (a) {[}(b){]}. The black
		lines separating different phases are obtained from the gapless condition
		Eq.~(\ref{eq:PhsBd}).\label{fig:PhsDiagJ1J2}}
\end{figure}

In the second case, we present the topological phase diagram of our
model versus the hopping amplitude $J_{1}$ and the imaginary part
of onsite potential $v$ in Fig.~\ref{fig:PhsDiagJ1V}. The other
system parameters are fixed at $J=\delta=\Delta/2=\pi/40$, $u=0$,
and $J_{2}=3\pi$. The values of topological invariants $\nu_{0}$
and $\nu_{\pi}$ for each of the phases are shown separately in the
panels (a) and (b) of Fig. \ref{fig:PhsDiagJ1V}, respectively. Similar
to the first case, we observe rich non-Hermitian Floquet SOTI phases
and phase transitions at different values of $J_{1}$ and $v$. Moreover,
around certain values of $J_{1}$ (e.g., $J_{1}=2.5\pi$), we find that by increasing the
gain and loss strength $v$, the system can shift to topological phases
with larger invariants, which could also support more quartets of
corner modes under the OBCs. Such kinds of non-Hermiticity \emph{enhanced} topological
properties are usually unexpected in systems with losses. Therefore,
it forms one of the defining features of our construction, with potential
applications in preparing Floquet topological states and combating environmental effects in quantum information tasks.

For completeness, we also present the phase diagram of our model versus the hopping amplitudes $J_1$ and $J_2$ in Fig.~\ref{fig:PhsDiagJ1J2}. It is clear that a series of topological phase transitions can be induced by varying both $J_1$ and $J_2$, yielding rich non-Hermitian Floquet SOTI phases. Furthermore, in certain ranges of $J_1$~(e.g., around $J_1=0.5\pi$), the magnitude of topological winding numbers $(\nu_0,\nu_\pi)$ tend to increase with $J_2$ monotonically. This observation again highlights the power of Floquet engineering in the realization of non-Hermitian SOTI phases with large topological invariants and multiple corner modes.

Note in passing that in the absence of the Floquet driving fields, our system could only possess a static non-Hermitian SOTI phase with winding number $\nu_0=1$, yielding at most four corner modes at zero energy under the OBCs. Thanks to the Floquet terms, the system could possess much richer SOTI phases with large topological winding numbers $(\nu_0,\nu_\pi)$, as presented by the phase diagrams. These phases are further subject to a ${\mathbb Z}\times{\mathbb Z}$ topological characterization, and therefore totally different from the static SOTI phases. As will be demonstrated in the next section, the non-Hermitian Floquet SOTI phases also possess many corner modes at both zero and $\pi$ quasienergies, with the $\pi$ modes being unique to Floquet systems. Therefore, the Floquet term is essential in generating the rich topological features of our system.

To summarize, we find rich non-Hermitian Floquet SOTI phases with
large topological invariants in our system. In the following two sections,
we discuss two experimentally relevant signatures of the intriguing
phases found in our system. We first present the Floquet spectrum and corner
modes of our system under the OBCs, and establish the correspondence between
the corner modes and the bulk topological invariants $(\nu_{0},\nu_{\pi})$.
Next, we show how to extract the invariants $(\nu_{0},\nu_{\pi})$
from the nonunitary stroboscopic dynamics of easily prepared wave packets.

\section{Corner states and bulk-corner correspondence\label{sec:CorStat}}

Under the OBCs, the Floquet operator of our periodically quenched
lattice model Eq.~(\ref{eq:Ukxy}) takes the form
\begin{equation}
U=U_{x}\otimes U_{y},\label{eq:UOBC}
\end{equation}
where
\begin{alignat}{1}
U_{x}& = e^{-{\rm i}\sum_{i,j}\frac{\Delta[1+(-1)^{i}]}{2}(|i,j\rangle\langle i+1,j|+{\rm H.c.})},\label{eq:UrOBC}\\
U_{y}& = e^{-{\rm i}\sum_{i,j}(-1)^{j}(iJ_{2}|i,j\rangle\langle i,j+2|-\mu|i,j\rangle\langle i,j|+{\rm H.c.})}\nonumber \\
& \times e^{-{\rm i}\sum_{i,j}J_{1}(|i,2j\rangle\langle i,2j+1|+{\rm H.c.})}.\label{eq:UcOBC}
\end{alignat}
The number of lattice sites along the $x$ ($y$) direction is $L_{x}=2N_{x}$
($L_{y}=2N_{y}$), with $N_{x}$ ($N_{y}$) being the number of unit
cells. The Floquet quasienergy spectrum and corner modes of the model
can be obtained by solving the eigenvalue equation $U|\Psi\rangle=e^{-iE}|\Psi\rangle$,
where $E$ is the quasienergy and $|\Psi\rangle$ is the corresponding
Floquet right eigenvector. Note that due to the balanced gain and loss in
the onsite potential $\mu=u+{\rm i}v$, the quasienergy $E$ is in
general a complex number. We define a quasienergy gap in this case
as a point on the complex plane, which is avoided by all the bulk
eigenstates for a given set of system parameters.

In the topological nontrivial regime, a 2D SOTI is featured by topologically
protected zero energy modes around the corners of the lattice. In
a non-Hermitian Floquet SOTI, there could be two types of topological
corner modes, whose quasienergies are zero and $\pi$. For the class
of periodically quenched lattice model studied in this work, the physical
origin of these corner modes can be directly inferred from the Kronecker
product structure of Floquet operator in Eq.~(\ref{eq:UOBC}) and its underlying sublattice symmetry ${\cal S}$. 
As discussed in Sec.~\ref{sec:Model}, our system can be viewed as an array of 1D Floquet topological insulators (FTIs) lying along the $y$-direction, with each of them being connected to its adjacent neighbors by SSH-type dimerized couplings along the $x$-direction. The number of zero and $\pi$ quasienergy edge modes of the 1D FTI is determined by its topological winding numbers $(w_0,w_\pi)=[(w_1+w_2)/2,(w_1-w_2)/2]$ following Eq.~(\ref{eq:w12}), whereas the number of zero-quasienergy edge modes of the 1D SSH chain is determined by its winding number $w$. When the 1D FTIs and 1D SSH chains are coupled to from our 2D Floquet system, there are only two possibilities for the localized modes at zero and $\pi$ quasienergies to appear. 
That is, if the 1D descendant systems $U_{x}$ and $U_{y}$ both possess
zero quasienergy edge modes, they will couple to form a Floquet corner
mode with quasienergy zero in the parent 2D system described by $U$ in Eq.~(\ref{eq:UOBC}), and the total number of these zero-quasienergy corner modes is determined by the invariant $\nu_0=ww_0$.
Similarly, if the 1D system $U_{x}$ ($U_{y}$) possesses a zero ($\pi$)
quasienergy edge mode, they will couple to form a Floquet corner mode
with quasienergy $\pi$ in the parent system $U=U_{x}\otimes U_{y}$, and the total number of these $\pi$ corner modes is determined by the invariant $\nu_\pi=ww_\pi$. The zero- and $\pi$-corner modes are robust to perturbations, so long as the sublattice symmetry ${\cal S}$ is preserved.
Moreover, the above analyses indicate that the number of non-Hermitian
Floquet corner modes with quasienergy zero ($\pi$) in the 2D system
is $n_{0}=n_{x0}n_{y0}$ ($n_{\pi}=n_{x0}n_{y\pi}$), where $n_{x0}$
is the number of zero edge modes of $U_{x}$ and $n_{y0}$ ($n_{y\pi}$)
is the number of zero ($\pi$) edge modes of $U_{y}$. Combining these
observations with the invariants $(\nu_{0},\nu_{\pi})$ defined in
Eq.~(\ref{eq:nu0P}), we could build the connection between the numbers
of non-Hermitian Floquet corner modes $(n_{0},n_{\pi})$ and the bulk
topological numbers as
\begin{equation}
n_{0}=4|\nu_{0}|,\qquad n_{\pi}=4|\nu_{\pi}|.\label{eq:BCC}
\end{equation}
Eq.~(\ref{eq:BCC}) establishes the bulk-corner correspondence of
2D chiral symmetric Floquet systems with the tensor product structure of Eq.~(\ref{eq:UOBC}), which also holds in the Hermitian limit ($\mu\in\mathbb{R}$) so long as
the sublattice symmetry ${\cal S}$ is retained.

Note in passing that
the system described by $U_{x}$ in Eq.~(\ref{eq:UrOBC}) is essentially
static, and therefore could only possess edge modes at zero quasienergy. If a zero (zero or $\pi$) edge mode of the SSH chain (1D FTI) is coupled to a bulk mode of the 1D FTI (SSH chain), it may result in an edge state with a finite quasienergy in the 2D system. Such kinds of finite-quasienergy (i.e., $E\neq0,\pi$) edge states are gapped, and their numbers will change with the system size as pointed out in Ref.~\cite{FHOTP1} for Hermitian systems. They are thus trivial gapped edge states. Therefore, there are no edge states at $E=0,\pi$, and the topological numbers in Eq.~(\ref{eq:nu0P}) only counts the number of zero and $\pi$ corner modes. The system is thus not a first-order topological system at $E=0,\pi$, although the topological numbers are well defined.

\begin{figure}
	\begin{centering}
		\includegraphics[scale=0.49]{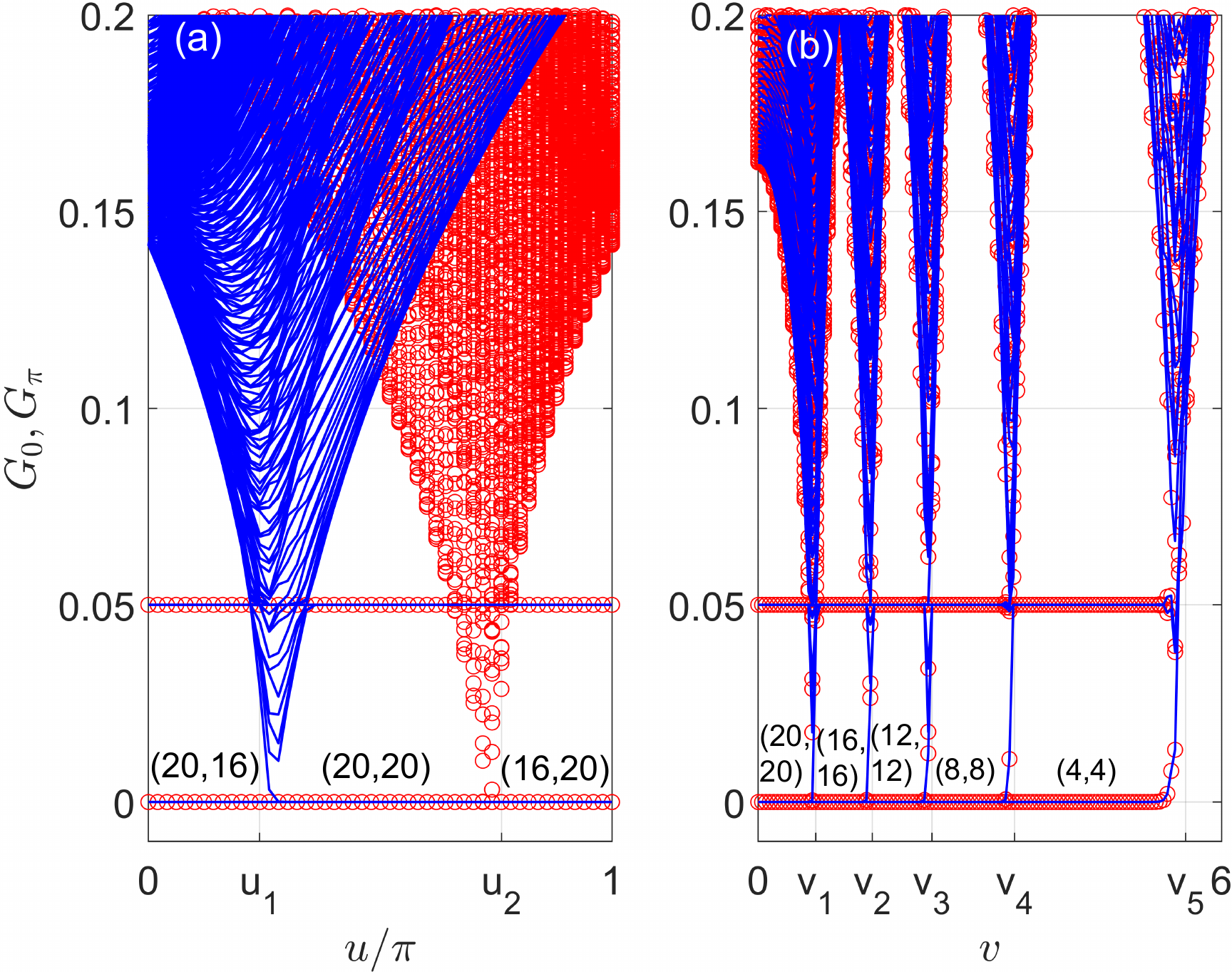}
		\par\end{centering}
	\caption{Spectral gap functions $G_{0}$ (red circles) and $G_{\pi}$ (blue
		lines) versus the real and imaginary parts of onsite potential $\mu=u+{\rm i}v$.
		The system parameters are $\Delta=\pi/20$, $J_{1}=0.5\pi$, $J_{2}=5\pi$
		and $v=0.5$ ($u=0.5\pi$) for panel (a) {[}(b){]}. The numbers of
		non-Hermitian Floquet topological corner modes at zero (with $G_{0}=0$)
		and $\pi$ (with $G_{\pi}=0$) quasienergies $(n_{0},n_{\pi})$ are
		denoted explicitly in each panel. The ticks $u_{i}$ ($i=1,2$)
		and $v_{i}$ ($i=1,2,3,4,5$) along the horizontal axis are the bulk
		phase transition points deduced from Eq.~(\ref{eq:PhsBd}).\label{fig:EvsUV}}
\end{figure}

To demonstrate the topological phase transitions and bulk-corner correspondence
of our system, we present the Floquet spectrum of $U$ in Eq.~(\ref{eq:UOBC})
under the OBCs for two typical examples. In order to show the evolution
of spectral gaps with the system parameters in a more transparent
manner, we introduce a pair of spectral gap functions, defined
as
\begin{alignat}{1}
G_{0}& = \frac{1}{\pi}\sqrt{({\rm Re}E)^{2}+({\rm Im}E)^{2}},\label{eq:G0}\\
G_{\pi}& = \frac{1}{\pi}\sqrt{(|{\rm Re}E|-\pi)^{2}+({\rm Im}E)^{2}}.\label{eq:GP}
\end{alignat}
It is clear that when the system becomes gapless at the quasienergy
$E=0$ ($E=\pm\pi$), we will have $G_{0}=0$ ($G_{\pi}=0$). In Fig.~\ref{fig:EvsUV}(a), we show the evolutions of $G_{0}$ (red circles)
and $G_{\pi}$ (blue lines) versus the real part $u$ of the onsite potential. The other system parameters are chosen as $\Delta=\pi/20$,
$J_{1}=0.5\pi$, $J_{2}=5\pi$ and $v=0.5$. The number of non-Hermitian
Floquet zero and $\pi$ corner modes $(n_{0},n_{\pi})$ are denoted
explicitly at $G_{0}=G_{\pi}=0$, and the ticks $(u_{1},u_{2})$
along the horizontal axis are determined by the gapless condition
Eq.~(\ref{eq:PhsBd}). We see that across each topological phase transition
point $u_{i}$ ($i=1,2$), the number of Floquet corner modes $n_{0}$
or $n_{\pi}$ changes by an integer multiple of four. Within the three
topological phases separated by $u_{1}$ and $u_{2}$, the numbers
of corner modes $(n_{0},n_{\pi})$ are related to the topological
invariants $(\nu_{0},\nu_{\pi})$ in Fig.~\ref{fig:PhsDiagUV} by
Eq.~(\ref{eq:BCC}), which confirms the bulk-corner correspondence
in our system. In Fig.~\ref{fig:EvsUV}(b), we show the changes of
$G_{0}$ (red circles) and $G_{\pi}$ (blue lines) versus the imaginary
part $v$ of the onsite potential, with the other system parameters
being set as $\Delta=\pi/20$, $J_{1}=0.5\pi$, $J_{2}=5\pi$ and
$u=0.5\pi$. Similarly, we observe that the numbers of corner modes
$(n_{0},n_{\pi})$ are related to the topological invariants $(\nu_{0},\nu_{\pi})$
by Eq.~(\ref{eq:BCC}). Moreover, the values of $(n_{0},n_{\pi})$ change by four
across every transition point $v_{i}$ ($i=1,2,3,4,5$) along the
horizontal axis. We thus conclude that a series of transitions between
different non-Hermitian Floquet SOTI phases can indeed be induced
by simply varying the magnitude of gain and loss rate $v$, and SOTI phases
unique to non-Hermitian Floquet systems could emerge after each transition
in our model.

\begin{figure}
	\begin{centering}
		\includegraphics[scale=0.49]{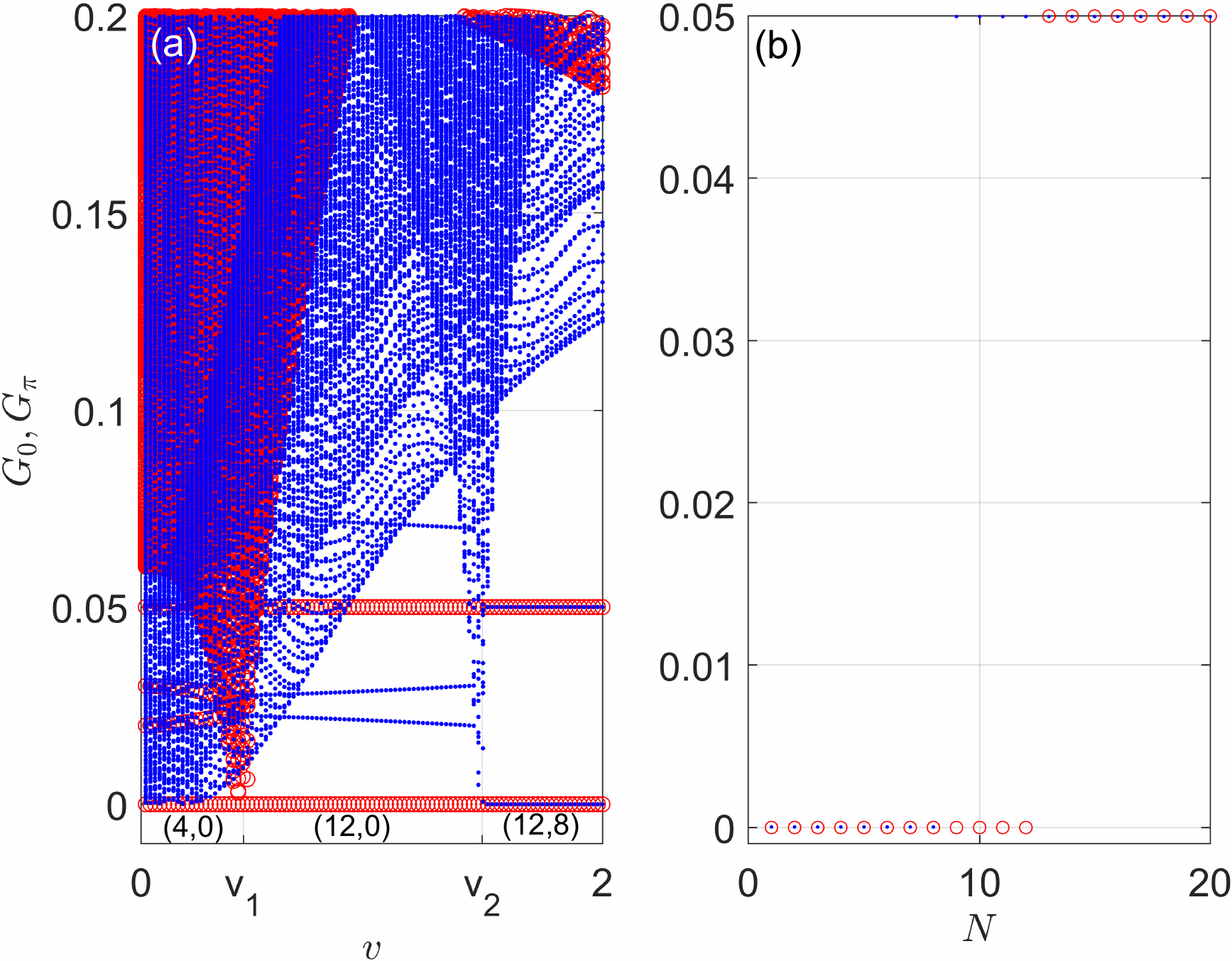}
		\par\end{centering}
	\caption{Spectral gap functions $G_{0}$ (red circles) and $G_{\pi}$ (blue
		dots) versus the imaginary part $v$ of the onsite potential in panel
		(a) and the state index $N$ in panel (b). The system parameters are
		set as $\Delta=\pi/20$, $J_{1}=2.5\pi$, $J_{2}=3\pi$ and $u=0$
		in panel (a), with further $v=2$ in panel (b). $v_{1}$ and $v_{2}$ refer
		to the topological phase transition points obtained from Eq.~(\ref{eq:PhsBd}),
		and the numbers of non-Hermitian Floquet zero and $\pi$ corner modes
		$(n_{0},n_{\pi})$ for each of the topological phases are denoted explicitly
		in panel (a).\label{fig:EvsVN}}
\end{figure}

\begin{figure}
	\begin{centering}
		\includegraphics[scale=0.48]{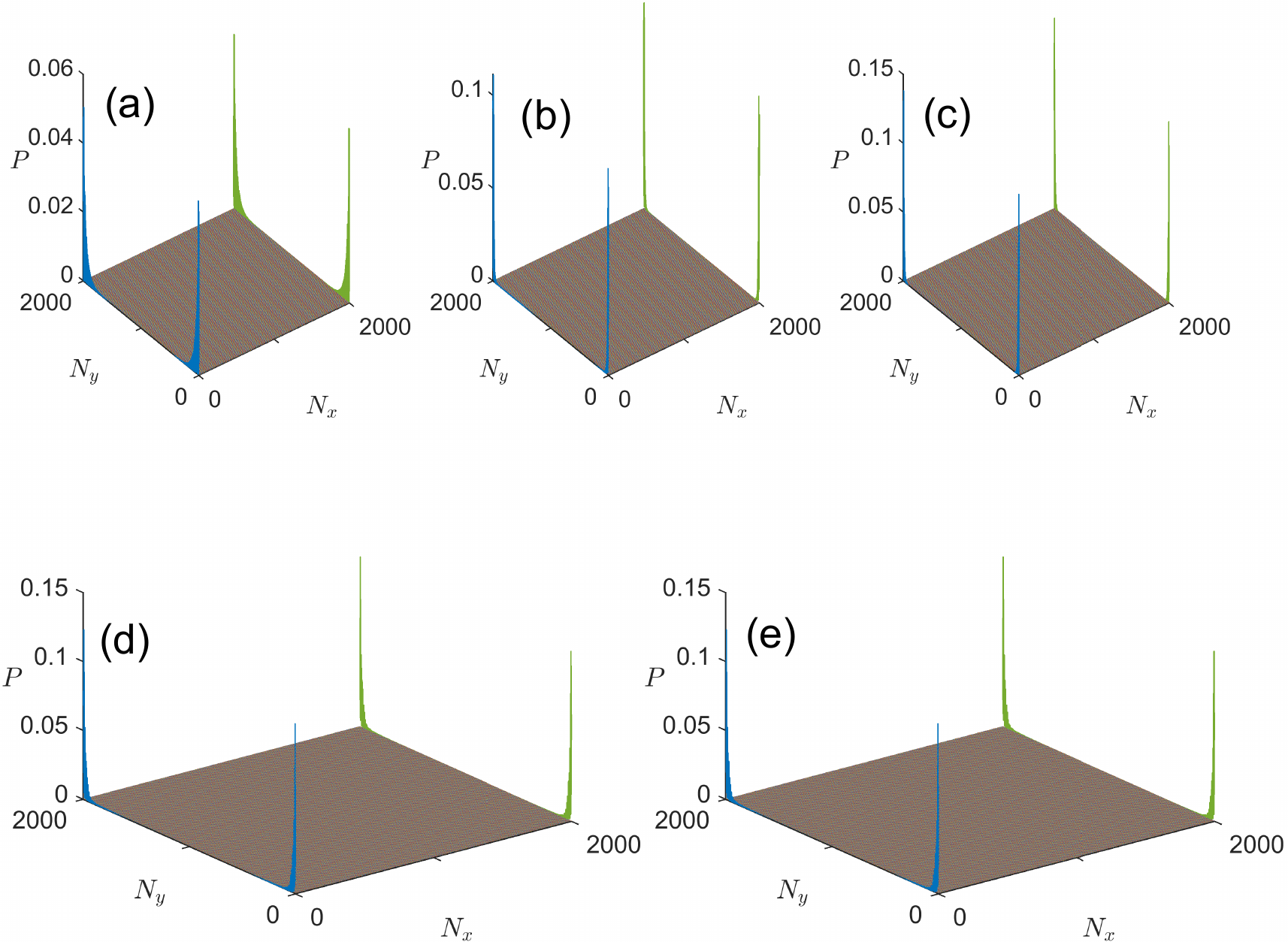}
		\par\end{centering}
	\caption{Probability distributions $P$ of the non-Hermitian Floquet topological
		corner modes at quasienergies zero {[}in panels (a), (b), (c){]} and
		$\pi$ {[}in panels (d), (e){]}, which are obtained from the right eigenvectors of the Floquet operator $U$ in Eq.~(\ref{eq:UOBC}). The lattice contains $N_{x}=N_{y}=2000$
		unit cells along $x$ and $y$ directions. The system parameters are
		the same as those of Fig.~\ref{fig:EvsVN}(b), yielding twelve zero
		corner modes and eight $\pi$ corner modes under the OBCs.\label{fig:Prob}}
\end{figure}

To further unveil the potential of non-Hermitian effects in generating
Floquet SOTI phases with more corner modes, we present the spectral
gap functions $(G_{0},G_{\pi})$ in red circles and blue dots versus
the gain and loss amplitude $v$ in Fig.~\ref{fig:EvsVN}(a). The
other system parameters are fixed at $\Delta=\pi/20$, $J_{1}=2.5\pi$,
$J_{2}=3\pi$ and $u=0$. It is clear that with the increase of $v$,
the numbers of corner modes $(n_{0},n_{\pi})$ changes from $(4,0)$
to $(12,0)$ across $v_{1}$ and from $(12,0)$ to $(12,8)$ across
$v_{2}$, coinciding with the bulk-corner relation Eq.~(\ref{eq:BCC}).
Such an \emph{enhancement} of topological signatures in deeper non-Hermitian
regimes is intriguing, which might be used to design new topological
state preparation schemes and achieve quantum information tasks in
open systems. To see the numbers and profiles of the Floquet corner
modes more explicitly, we show the first twenty states of the system
at $v=2$ in Fig. \ref{fig:EvsVN}(b), with the other system parameters
chosen to be the same as in Fig. \ref{fig:EvsVN}(a). The twelve (eight)
non-Hermitian Floquet corner modes at the quasienergy zero $(\pi)$
are denoted by red circles (blue dots), whose probability distributions
are shown in Figs.~\ref{fig:Prob}(a)-(c) {[}Figs.~\ref{fig:Prob}(d)-(e){]}. 
Here we plotted the distributions of right eigenvectors of $U$ in Eq.~(\ref{eq:UOBC}) in the lattice representation, and similar results can be obtained from the left eigenvectors.
We see that the zero and $\pi$
modes are indeed well-localized around the four corners of the 2D
lattice, which are protected by the sublattice symmetry ${\cal S}=\sigma_z\otimes\tau_y$
introduced in Sec.~\ref{sec:Model}.

\begin{figure}
	\begin{centering}
		\includegraphics[scale=0.48]{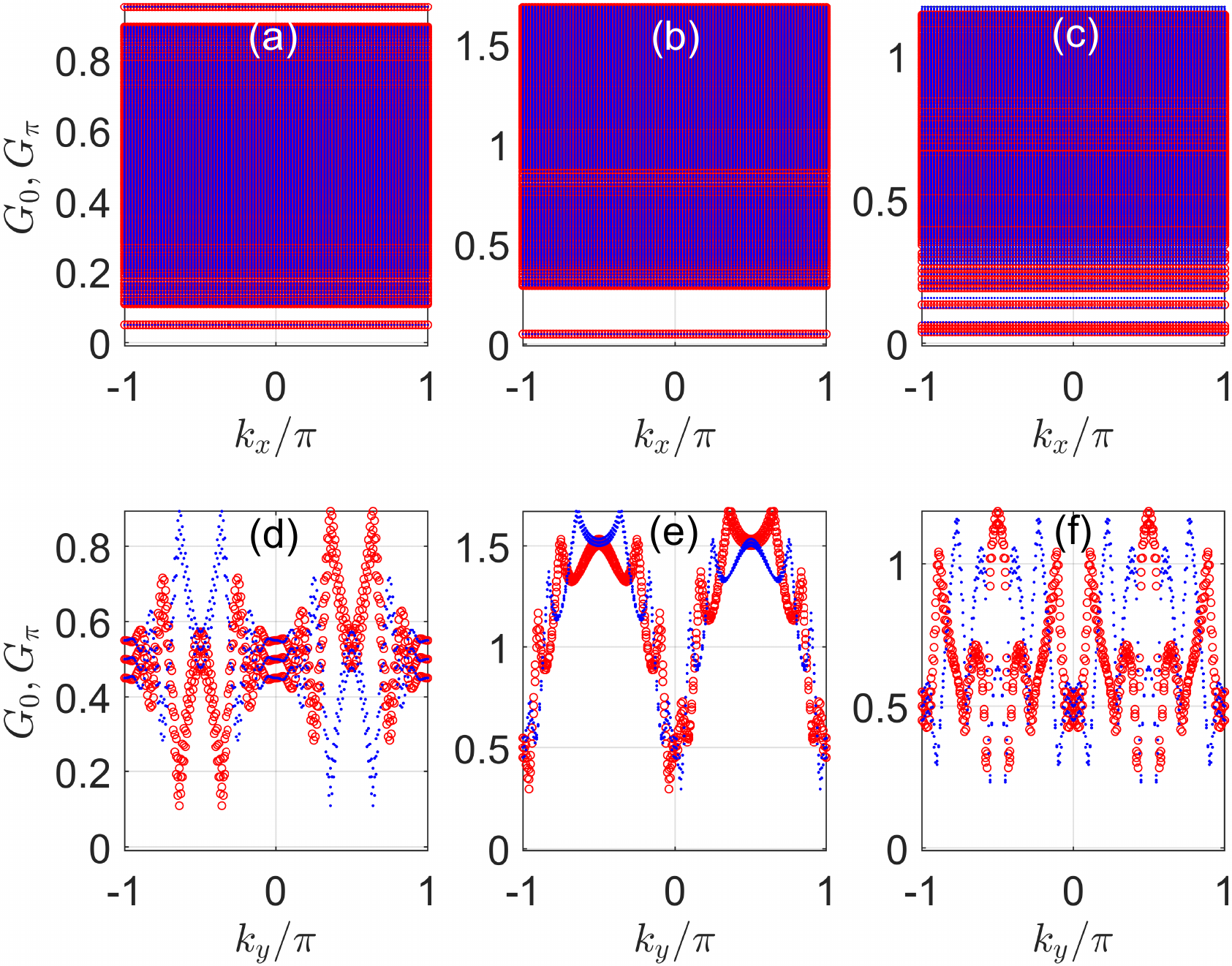}
		\par\end{centering}
	\caption{Spectral gap functions $G_0$ (in red circles) and $G_\pi$ (in blue dots) versus the quasimomentum $k_x$ ($k_y$) under the PBC (OBC) along $x$-direction and OBC (PBC) along $y$-direction of the lattice, respectively, in panels (a)-(c) [(d)-(f)]. The system parameters are chosen as $(J_1,J_2,\mu,\Delta)=(0.5\pi,5\pi,0.5\pi+0.5{\rm i},0.05\pi)$ for panels (a), (d); $(J_1,J_2,\mu,\Delta)=(0.5\pi,5\pi,0.5\pi+4.5{\rm i},0.05\pi)$ for panels (b), (e); and $(J_1,J_2,\mu,\Delta)=(2.5\pi,3\pi,2{\rm i},0.05\pi)$ for panels (c), (f). The complex Floquet spectrum is gapped around zero and $\pi$ quasienergies for all cases.\label{fig:EMBC}}
\end{figure}

For completeness, in Fig.~\ref{fig:EMBC} we present the gap functions with respect to the quasimomentum $k_x$ ($k_y$) by taking the PBC (OBC) along $x$-direction and OBC (PBC) along $y$-direction of the lattice. For all the three cases considered in Fig.~\ref{fig:EMBC}, the systems are set in non-Hermitian Floquet SOTI phases, and the gap functions $G_0$ and $G_\pi$ are found to be gapped at $G_0=G_\pi=0$. This means that in the complex Floquet spectrum of the system, all possible 1D edge states are gapped at the quasienergies zero and $\pi$, as expected for SOTI phases.


\section{Dynamical characterization of the topological phases\label{sec:MCD}}

The mean chiral displacement (MCD) is first introduced as the time-averaged
chiral displacement $\hat{x}{\cal S}$ of an initially localized wave
packet in a 1D lattice within the symmetry classes AIII and BDI~\cite{MCD1}.
Later, it is generalized to Floquet systems~\cite{LWZH1,MCD2}, non-Hermitian systems~\cite{LWZNH2,LWZNH3},
interacting systems~\cite{MCD3}, systems in other symmetry classes~\cite{LWZH3} and higher
physical dimensions~\cite{FHOTP1}. Experimentally, the MCD has been measured in
cold atom~\cite{MCD4} and photonic~\cite{MCD5} setups. In this section, we extend the definition
of MCD to 2D non-Hermitian Floquet systems with sublattice symmetry,
and demonstrate how to extract the topological invariants of our model
dynamically from the MCDs.

For a 2D lattice model with the the sublattice symmetry ${\cal S}$,
we define its chiral displacement operator as ${\cal C}=\hat{r}\otimes{\cal S}$,
with $\hat{r}$ being the unit cell position operator. The chiral
displacement of a wave packet $\rho_{0}$ in the symmetric time frame
$\alpha$ is then given by
\begin{equation}
C_{\alpha}(t)={\rm Tr}\left(\rho_{0}\tilde{U}_{\alpha}^{\dagger t}{\cal C}U_{\alpha}^{t}\right),\label{eq:Ca}
\end{equation}
where $\alpha=1,2$, $t$ counts the number of driving periods, and the
trace ${\rm Tr}(\cdots)$ is taken over all degrees of freedom of the system. To build
the connection between $C_{\alpha}$ and the topological invariants $(\nu_0,\nu_{\pi})$
of the system in the most straightforward manner, we prepare the initial
state $\rho_{0}$ in the central unit cell $(m,n)=(0,0)$ of the lattice,
with all the four sublattices being uniformly filled, i.e., $\rho_{0}=|0,0\rangle\langle0,0|\otimes\mathbb{I}_{4}/4$.
For our periodically quenched lattice model Eq.~(\ref{eq:Ukxy}), $U_{\alpha}$
is given by the inverse Fourier transform of Eq.~(\ref{eq:Uakxy}),
and $\tilde{U}_{\alpha}$ is defined such that if $|\Psi\rangle$
is a right eigenvector of $U_{\alpha}$ with eigenvalue $e^{-iE}$,
it is the left eigenvector of $\tilde{U}_{\alpha}$ with the same
eigenvalue. 

In the following, we will relate the long-time average of $C_{\alpha}(t)$
to the topological invariants of 2D non-Hermitian Floquet operators
with the structure of Eq.~(\ref{eq:Uakxy}) and the sublattice symmetry
${\cal S}$. Note that in the Hermitian limit, we simply have $\tilde{U}_{\alpha}=U_{\alpha}$,
and our derivations below will also hold. Taking the trace in Eq.~(\ref{eq:Ca}) explicitly and inserting the identities in the lattice and momentum representations, we find
\begin{alignat}{1}
&C_{\alpha}(t) = \frac{1}{4}\sum_{k_{x},k_{y},k'_{x},k'_{y}}\sum_{mn}mn\langle0,0|k_{x},k_{y}\rangle\langle k_{x},k_{y}|m,n\rangle\nonumber \\
\times &\langle m,n|k'_{x},k'_{y}\rangle\langle k'_{x},k'_{y}|0,0\rangle{\rm tr}\left[\tilde{U}_{\alpha}^{\dagger t}(k_{x},k_{y}){\cal S}U_{\alpha}^{t}(k'_{x},k'_{y})\right],\label{eq:Cat0}
\end{alignat}
where the trace ${\rm tr}[\cdots]$ in the second line is only taken over
the sublattice degrees of freedom. Using the Fourier expansion $|m,n\rangle=\frac{1}{\sqrt{N_{x}N_{y}}}\sum_{k_{x},k_{y}}e^{{\rm i}(k_{x}m+k_{y}n)}|k_{x},k_{y}\rangle$,
Eq.~(\ref{eq:Cat0}) can be simplified to
\begin{alignat}{1}
C_{\alpha}(t)& =  \frac{1}{4}\sum_{k_{x},k'_{x}}\sum_{k_{y},k'_{y}}\sum_{mn}\frac{mn}{N_{x}^{2}N_{y}^{2}}e^{{\rm i}(k_{x}m+k_{y}n)}\nonumber \\
& \times e^{-{\rm i}(k'_{x}m+k'_{y}n)}{\rm tr}\left[\tilde{U}_{\alpha}^{\dagger t}(k_{x},k_{y}){\cal S}U_{\alpha}^{t}(k'_{x},k'_{y})\right].\label{eq:Cat1}
\end{alignat}
With the help of summation formulas $\sum_{m}me^{{\rm i}(k_{x}-k'_{x})m}={\rm i}N_{x}\partial_{k'_{x}}\delta_{k_{x}k'_{x}}$
and $\sum_{n}ne^{{\rm i}(k_{y}-k'_{y})n}={\rm i}N_{y}\partial_{k'_{y}}\delta_{k_{y}k'_{y}}$,
we further obtain
\begin{alignat}{1}
C_{\alpha}(t)& = \frac{1}{4}\sum_{k_{x},k'_{x}}\sum_{k_{y},k'_{y}}\frac{1}{N_{x}N_{y}}({\rm i}\partial_{k'_{x}}\delta_{k_{x}k'_{x}})\nonumber \\
& \times ({\rm i}\partial_{k'_{y}}\delta_{k_{y}k'_{y}}){\rm tr}\left[\tilde{U}_{\alpha}^{\dagger t}(k_{x},k_{y}){\cal S}U_{\alpha}^{t}(k'_{x},k'_{y})\right].\label{eq:Cat2}
\end{alignat}
Finally, taking the continuous limit $N_{j}\rightarrow\infty$,
we have $N_{j}\delta_{k_{j}k'_{j}}\rightarrow\delta(k_{j}-k'_{j})$
and $\sum_{k_{j},k'_{j}}\rightarrow N_{j}^{2}\int_{-\pi}^{\pi}\frac{dk_{j}}{2\pi}\int_{-\pi}^{\pi}\frac{dk'_{j}}{2\pi}$
for $j=x,y$. Eq.~(\ref{eq:Cat2}) then becomes
\begin{alignat}{1}
C_{\alpha}(t)& =\int_{-\pi}^{\pi}\frac{dk_{x}}{2\pi}\int_{-\pi}^{\pi}\frac{dk_{y}}{2\pi}\nonumber \\
& \times \frac{1}{4}{\rm tr}\left[\tilde{U}_{\alpha}^{\dagger t}(k_{x},k_{y}){\cal S}({\rm i}\partial_{k_{x}}{\rm i}\partial_{k_{y}})U_{\alpha}^{t}(k{}_{x},k{}_{y})\right].\label{eq:Cat3}
\end{alignat}
For our periodically quenched lattice model, the expression of chiral
displacement $C_{\alpha}(t)$ can be further simplified. Noting the tensor
product structure of Floquet operator $U_{\alpha}$ in Eq.~(\ref{eq:Uakxy})
and the expression of sublattice symmetry operator ${\cal S}=\sigma_{z}\otimes\tau_{y}$,
we can write $C_{\alpha}(t)$ as a product of chiral displacements in the
descendant 1D systems as $C_{\alpha}(t)=C_{x}(t)C_{\alpha y}(t)$~\cite{Note1},
where
\begin{alignat}{1}
C_{x}(t)= & \int_{-\pi}^{\pi}\frac{dk_{x}}{2\pi}\frac{1}{2}{\rm tr}\left[{\cal U}_{0}^{\dagger t}(k_{x})\sigma_{z}{\rm i}\partial_{k_{x}}{\cal U}_{0}^{t}(k_{x})\right],\label{eq:Cxt}\\
C_{\alpha y}(t)= & \int_{-\pi}^{\pi}\frac{dk_{y}}{2\pi}\frac{1}{2}{\rm tr}\left[\tilde{{\cal U}}_{\alpha}^{\dagger t}(k_{y})\tau_{y}{\rm i}\partial_{k_{y}}{\cal U}_{\alpha}^{t}(k_{y})\right].\label{eq:Cayt}
\end{alignat}
Summing up the chiral displacements $C_{\alpha}(t)$ over different numbers $t$ of
the driving period and taking the long-time average, we obtain the
MCD of our system in the $\alpha$'s time frame as
\begin{alignat}{1}
\overline{C}_{\alpha}= & \lim_{t\rightarrow\infty}\frac{1}{t}\sum_{t'=1}^{t}C_{x}(t')\label{eq:Cbara}\\
\times & \int_{-\pi}^{\pi}\frac{dk_{y}}{2\pi}\frac{1}{2}\frac{{\rm tr}\left[\tilde{{\cal U}}_{\alpha}^{\dagger t'}(k_{y})\tau_{y}i\partial_{k_{y}}{\cal U}_{\alpha}^{t'}(k_{y})\right]}{{\rm tr}\left[\tilde{{\cal U}}_{\alpha}^{\dagger t'}(k_{y}){\cal U}_{\alpha}^{t'}(k_{y})\right]},\nonumber 
\end{alignat}
where we have inserted a normalization factor ${\rm tr}\left[\tilde{U}_{\alpha}^{\dagger t'}(k_{x},k_{y})U_{\alpha}^{t'}(k{}_{x},k{}_{y})\right]={\rm tr}\left[\tilde{{\cal U}}_{\alpha}^{\dagger t'}(k_{y}){\cal U}_{\alpha}^{t'}(k_{y})\right]$
to compensate for the changing norm of the state
during the nonunitary evolution. Note that the same expression for
$\overline{C}_{\alpha}$ can be derived if the dynamics is expressed
in the biorthogonal basis~\cite{BiQM1}.

In previous studies, it has been shown that under the limit $\lim_{t\rightarrow\infty}\frac{1}{t}\sum_{t'=1}^{t}$,
$C_{x}(t')$ is averaged to $w/2$~\cite{LWZH1} and the second line in Eq.~(\ref{eq:Cbara}) converges to $w_{\alpha}/2$~\cite{LWZNH2}. Putting together,
we would obtain 
\begin{equation}
\overline{C}_{\alpha}=ww_{\alpha}/4=\nu_{\alpha}/4\label{eq:Cbara0}
\end{equation}
for $\alpha=1,2$ according to Eq.~(\ref{eq:nu12}). Therefore, with
the help of Eq.~(\ref{eq:nu0P}), we establish the connection between
the topological invariants $(\nu_{0},\nu_{\pi})$ and the MCDs as
\begin{alignat}{1}
\nu_{0}= & 2(\overline{C}_{1}+\overline{C}_{2})\equiv2C_{0},\label{eq:MCD0}\\
\nu_{\pi}= & 2(\overline{C}_{1}-\overline{C}_{2})\equiv2C_{\pi}.\label{eq:MCDP}
\end{alignat}
These relations have been derived before for Hermitian Floquet SOTIs~\cite{FHOTP1}. Upon appropriate modifications, we find that they also
hold in non-Hermitian Floquet systems with the sublattice symmetry
${\cal S}$. Experimentally, by measuring the MCDs $(\overline{C}_{1},\overline{C}_{2})$
of the dynamics over a long time-duration, we would be able to extract
the topological invariants $(\nu_{0},\nu_{\pi})$ for the class of
non-Hermitian Floquet SOTI models studied in this work.

\begin{figure}
	\begin{centering}
		\includegraphics[scale=0.5]{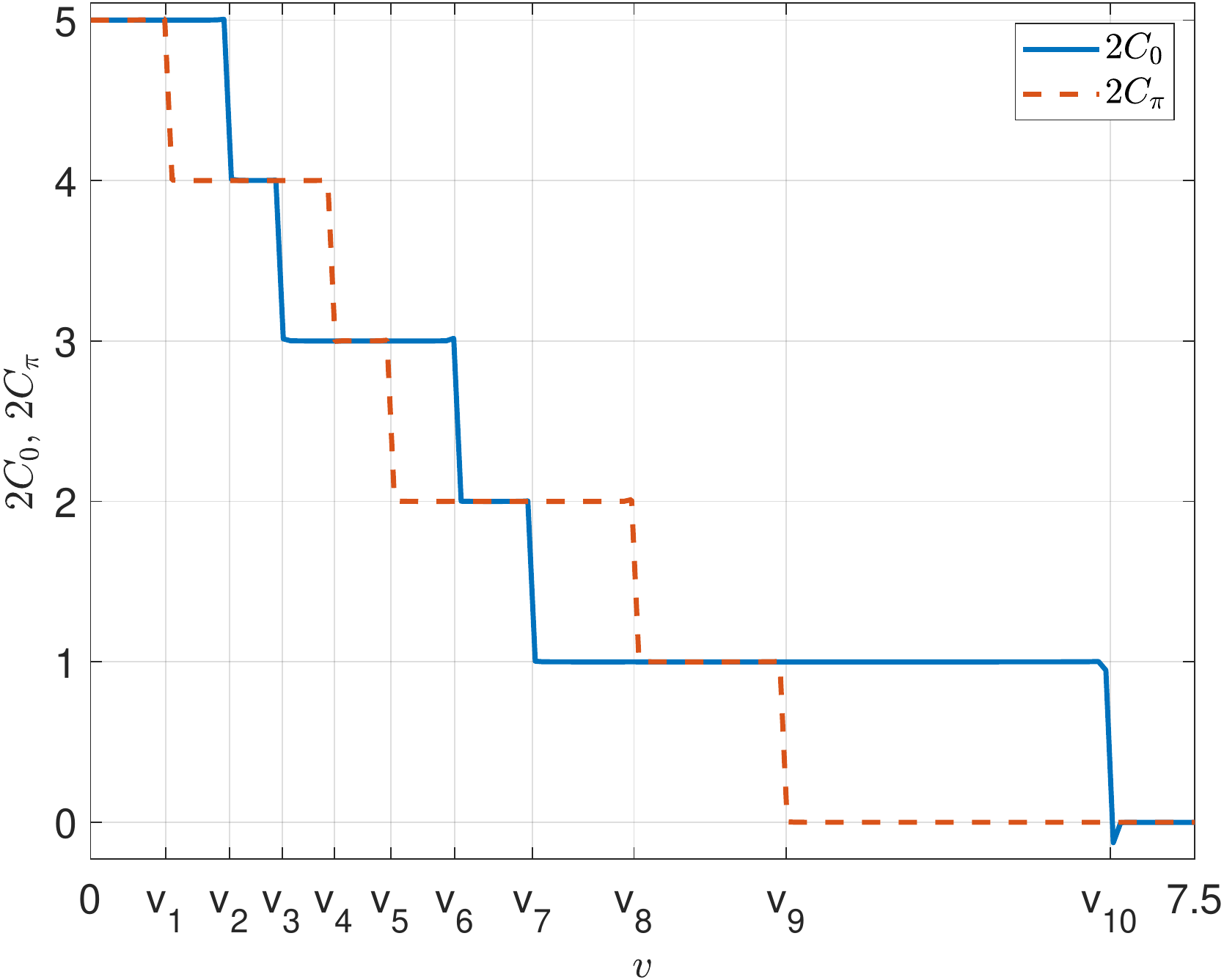}
		\par\end{centering}
	\caption{The MCDs $(C_{0},C_{\pi})$ versus the imaginary part of onsite potential
		$v$, after averaging over $M=100$ driving periods. The other system
		parameters are chosen as $\Delta=\pi/20$, $J_{1}=0.5\pi$, $J_{2}=5\pi$
		and $u=0.25\pi$. The transition points $v_{i}$ for $i=1,...,10$,
		separating different non-Hermitian Floquet SOTI phases, are extracted
		from Eq.~(\ref{eq:PhsBd}). The gray horizontal grids are guiding lines
		of topological invariants $(\nu_{0},\nu_{\pi})$ in each phase, whose
		values are related to the MCDs $(C_{0},C_{\pi})$ through the Eqs.~(\ref{eq:MCD0}) and (\ref{eq:MCDP}).\label{fig:MCDs}}
\end{figure}

To be concrete, we present a typical example of the recombined MCDs
$(C_{0},C_{\pi})$ obtained numerically from Eq.~(\ref{eq:Cbara})
for our periodically quenched lattice model Eq.~(\ref{eq:Ukxy}) in
Fig.~\ref{fig:MCDs}. The system parameters are chosen to be $\Delta=\pi/20$,
$J_{1}=0.5\pi$, $J_{2}=5\pi$, $u=0.25\pi$, and the dynamics is
averaged over $M=100$ driving periods. From Fig.~\ref{fig:MCDs},
we see clearly that the value of $C_{0}$ or $C_{\pi}$ gets a quantized
jump every time when the imaginary part $v$ of the onsite potential reaches
a topological phase transition point $v_{i}$ ($i=1,...,10$), as
predicted by Eq.~(\ref{eq:PhsBd}). Furthermore, between each pair
of adjacent transition points, the values of $(2C_{0},2C_{\pi})$
remain quantized, equaling to the topological invariants $(\nu_{0},\nu_{\pi})$
of the corresponding non-Hermitian Floquet SOTI phase as shown in
Fig.~\ref{fig:PhsDiagUV}. Putting together, we verified the correctness
of the relations in Eqs.~(\ref{eq:MCD0}) and (\ref{eq:MCDP}) between the bulk
topological invariants and MCDs of non-Hermitian Floquet SOTIs with
sublattice symmetry. In the meantime, these results demonstrate the
usefulness of MCDs in characterizing and detecting topological phases
and phase transitions in 2D non-Hermitian Floquet systems. Numerically,
we observe good quantizations of $(2C_{0},2C_{\pi})$ for an average
over as few as $M=15$ driving periods, which should be well within
reach under current experimental conditions.

In experiments, the MCD could been detected in both cold atom and photonic systems. In a photonic setup, the MCD could be obtained from the the quantum walk of twisted photons by measuring the Zak phase~\cite{MCD1}, or from the chiral intensity distribution of structured light~\cite{MCD3}. In a cold atom setup, the MCD can be obtained from the time-of-flight images at different time steps of the evolution of a wave packet, which is initially prepared at central unit cell of the lattice and then subjected to periodically switched lattice parameters~\cite{MCD4,MCD5}. Since our system can be viewed as the Kronecker sum of two 1D systems, and the non-Hermitian term can be engineered in both cold atom and photonic systems, we expect that the MCDs we introduced could be detectable in both cold atom and photonic setups.

\section{Summary and discussion\label{sec:Summary}}

In this work, we found rich non-Hermitian Floquet SOTI phases in
periodically quenched 2D lattices with balanced gain and loss. Each
of the phases is characterized by a pair of integer topological invariants
$\nu_{0}$ and $\nu_{\pi}$, which allow us to establish the topological
phase diagram of the model. We further observed multiple non-Hermitian
Floquet SOTI phases with large topological invariants and various
gain or loss-induced topological phase transitions. Under the OBCs, the
invariants $\nu_{0}$ and $\nu_{\pi}$ predict the numbers of protected
Floquet corner modes at the quasienergies zero and $\pi$. Thanks
to the interplay between the periodic drivings and non-Hermitian effects,
we found a series of non-Hermitian Floquet SOTI phases with many zero
and $\pi$ corner modes, which might be useful in topological state preparations, detections and quantum information technologies. Finally, we introduced a generalized version of the mean
chiral displacement, which could capture the topological invariants
of our system through the wave packet dynamics.

Before discussing the experimental realization of our model and possible future directions, the essential role played by the non-Hermitian term in our system deserve to be emphasized. First, a series of topological phase transitions can be induced by varying the non-Hermitian term as reflected in the phase diagrams, and rich non-Hermitian Floquet SOTI phases could emerge after these transitions. These new phases could persist only when the system is subject to both the driving fields and the non-Hermitian effects. Therefore, they are \textit{unique} to non-Hermitian Floquet systems, different from any phases that may appear in the system if the non-Hermitian term is switched off. Second, in the phase diagram with respect to $J_1$ and $v$, we observe that with the increase of the non-Hermitian term $v$ around $J_1=2.5\pi$, the system can undergo a transition from a non-Hermitian Floquet SOTI phase with winding numbers $(\nu_0,\nu_\pi)=(1,0)$ to another phase with $(\nu_0,\nu_\pi)=(3,-2)$. This means that the resulting phase could carry larger topological invariants and more topological corner modes when the gains and losses become stronger, which clearly runs counter to the belief that the non-Hermitian term is usually destructive for topological phases. The underlying physics behind this intriguing observation is again the interplay between the losses and driving fields, for which the non-Hermitian term is necessary. Putting together, the SOTI phases discovered in our system are different from those in static systems, in the sense that the former and later are characterized by distinct topological invariants and phase transitions. They are also different from SOTI phases in Hermitian Floquet systems, as the non-Hermitian term could create new phase transitions and SOTI phases with even larger topological invariants compared with the Hermitian counterparts. Our work thus extend the study of SOTIs to physical settings with both drivings, gains	and losses, and unraveled the richness of non-Hermitian Floquet SOTI phases that can appear in such situations.

A candidate setup in which the bulk Floquet operator of our system might be realizable is the nitrogen-vacancy-center in diamond~\cite{LWZH2,NVExp1}. By applying a universal dilation scheme, an arbitrary non-Hermitian model with a finite number of bands can in principle be mapped to a Hermitian Hamiltonian in an enlarged Hilbert space~\cite{NVExp1}. The non-Hermitian Floquet band structure and dynamics of our system can then be studied with the help of the dilated Hamiltonian and its resulting unitary evolution, in which the periodic driving can also be implemented~\cite{LWZH2}. In the definition of our model, we have set the driving period $T=1$, leading to a dimensionless driving frequency $\omega=2\pi$. The other system parameters used in the phase diagrams are either smaller then or comparable to $\omega$. According to the setups introduced in~\cite{LWZH2,NVExp1}, we expect that the choices of system parameters in our model should be within reach under current or near-term experimental conditions. Another possible setup that could be used to realize our model is the cold atom system. In cold atom systems, there are mature technologies to realize topological bands in different physical dimensions~\cite{CdAtmRev1,CdAtmRev2}. An SOTI might then be realized by loading ultracold atoms into the orbital angular momentum states of an optical lattice~\cite{CdAtmSOTI}. The non-Hermitian term in our system might be engineered by staggered onsite atom losses. To obtain such losses, one could introduce resonant couplings between the ground and excited states of atoms, which realizes the effective loss for the ground state and also controls the staggered loss~\cite{NHHOTP3}. The staggered loss is further equivalent to the staggered gain-loss configuration in our system up to a constant. Finally, the periodic quenches can be achieved by stepwise Raman-induced couplings~\cite{MCD5}. In the cold atom setup realized by Ref.~\cite{MCD5}, the magnitude of the hopping rate is $\hbar\Omega$, where the Raman-coupling rate $\Omega\sim2\pi\times2.3$kHz. The driving period realized in the experiment is around $0.22$ms, which corresponds to a driving frequency $\omega\sim2\pi\times4.5$kHz. From these experimental data, it is clear that the realized Floquet hopping amplitude $\hbar\Omega$ and the energy scale of driving photon $\hbar\omega$ are comparable. On the other hand, the driving frequency of our model is $\omega=2\pi$ in dimensionless units, and the Floquet hopping terms $J_1$ and $J_2$ are set within the range of $(0,3\pi)$ for most of our numerical examples. Therefore, referring to the experiment performed in Ref.~\cite{MCD5}, the system parameters involved in our numerical siumlations are expected to be reasonable, as their magnitudes are either smaller then or comparable to the (dimensionless) driving frequency $\omega=2\pi$. Putting together, we expect that our model should be realizable in cold atom systems as well in the context of current or near-term experimental technologies.

In future work, it would be interesting to generalize our strategies
to the engineering of non-Hermitian Floquet HOTPs in other symmetries
classes and higher spatial dimensions. 
For example, due to the sublattice symmetry ${\cal S}$, the spatial symmetries ${\cal M}_{\pm}$ and the configuration of hopping amplitudes $J_{1,2}$, the corner modes of our model are expected to appear at the four corners of a lattice with a square-shaped boundary. In systems with honeycomb or kagome lattice structures, the SOTI phases would be protected by a different set of crystal and rotational symmetries, and the corner modes might be observable under a triangular-shaped boundary~\cite{Classify3}. Finding non-Hermitian Floquet SOTI phases in such kinds of lattices would be an interesting topic for further study.
Moreover, in superconducting
systems, the interplay between drivings and non-Hermitian effects
may also induce multiple quartets of Floquet Majorana corner modes,
which are potentially useful in realizing certain topological quantum computing
tasks~\cite{FHOTP2,FHOTP11}.

\vspace{0.5cm}

\section*{Acknowledgement}
L. Z. is supported by the National Natural Science Foundation of China (Grant No.~11905211), the China Postdoctoral Science Foundation (Grant No.~2019M662444), the Fundamental Research Funds for the Central Universities (Grant No.~841912009), the Young Talents Project at Ocean University of China (Grant No.~861801013196), and the Applied Research Project of Postdoctoral Fellows in Qingdao (Grant No.~861905040009).



\end{document}